%% file: main.tex
\documentclass[ALICE,manyauthors]{cernphprep}
\usepackage[comma,square,numbers,sort&compress]{natbib}
\usepackage{hyperref}
\usepackage{lineno}
\usepackage{xspace}
\usepackage{color}
\usepackage{amsmath, amssymb}
\usepackage[T1]{fontenc}
\usepackage{orcidlink}

\begin{document}
\input{commands.tex}

\begin{titlepage}
\PHyear{2024}       
\PHnumber{133}      
\PHdate{21 May}  

\title{Investigating strangeness enhancement with multiplicity in pp collisions using angular correlations}
\ShortTitle{Strangeness enhancement in pp collisions using angular correlations}   

\Collaboration{ALICE Collaboration\thanks{See Appendix~\ref{app:collab} for the list of collaboration members}}
\ShortAuthor{ALICE Collaboration} 

\begin{abstract}
A study of strange hadron production associated with hard scattering processes and with the underlying event is conducted to investigate the origin of the enhanced production of strange hadrons in small collision systems characterised by large charged-particle multiplicities.
For this purpose, the production of the single-strange meson ${\rm K}^{0}_{\rm{S}}$ and the double-strange baryon $\Xi^{\pm}$ is measured, in each event, in the azimuthal direction of the highest-$p_{\rm T}$ particle (``trigger" particle), related to hard scattering processes, and in the direction transverse to it in azimuth, associated with the underlying event, in pp collisions at $\sqrt{s}=5.02$~TeV and $\sqrt{s}=13$~TeV using the ALICE detector at the LHC. 
The \mbox{per-trigger} yields of ${\rm K}^{0}_{\rm{S}}$ and $\Xi^{\pm}$ are dominated by the transverse-to-leading production (i.e., in the direction transverse to the trigger particle), whose contribution relative to the toward-leading production is observed to increase with the event charged-particle multiplicity.
The transverse-to-leading and the toward-leading $\Xi^{\pm}$/${\rm K}^{0}_{\rm{S}}$ yield ratios increase with the multiplicity of charged particles, suggesting that strangeness enhancement with multiplicity is associated with both hard scattering processes and the underlying event.
 The relative production of $\Xi^{\pm}$ with respect to ${\rm K}^{0}_{\rm{S}}$ is higher in transverse-to-leading processes over the whole multiplicity interval covered by the measurement.
The ${\rm K}^{0}_{\rm{S}}$ and $\Xi^{\pm}$ \mbox{per-trigger} yields and yield ratios are compared with predictions of three different phenomenological models, namely \textsc{Pythia}8.2 with the Monash tune, \textsc{Pythia}8.2 with ropes and EPOS LHC. The comparison shows that none of them can quantitatively describe either the transverse-to-leading or the toward-leading yields of ${\rm K}^{0}_{\rm{S}}$ and $\Xi^{\pm}$.
\end{abstract}
\end{titlepage}

\setcounter{page}{2} 


\section{Introduction} 
The enhancement of strange hadron production in heavy-ion collisions with respect to minimum bias pp collisions was one of the first predicted signatures of quark--gluon plasma (QGP) formation~\cite{PhysRevLett.48.1066, KOCH1983151, KOCH1986167}. 
This strangeness enhancement was first observed at the SPS~\cite{Bartke1990, WA97, Antinori:2006ij, Alt:2004kq, PhysRevC.78.034918, PhysRevC.80.034906} and was later measured in Au--Au collisions at RHIC~\cite{PhysRevC.77.044908} and in Pb--Pb collisions at the LHC~\cite{ALICE:2013xmt}. The ALICE Collaboration further studied the production of strange hadrons in smaller collision systems, such as p--Pb~\cite{ALICE:2015mpp, ALICE:2013wgn, CMS:2019isl} and pp collisions~\mbox{\cite{ALICE:2016fzo, ALICE:2018pal, ALICE:2020jsh, ALICE:2019avo, LHCb:2010rlm, CMS:2011jlm, ATLAS:2011xhu}}. The results show that the ratios of (multi-)strange to non-strange hadron yields increase with the multiplicity of charged particles produced in the collision, reaching in high-multiplicity pp collisions values compatible with those measured in peripheral Pb--Pb collisions~\cite{ALICE:2020nkc}. 

The smooth evolution of the ratios with multiplicity across different collision systems implies a common particle production mechanism in the different systems. This is also supported by other observables, such as ``ridge''-like structures in the two-particle angular correlations at large pseudorapidity difference~\cite{ALICE:2021nir, PhysRevLett.116.172302} and non-vanishing anisotropic flow coefficients~\cite{CMS:2016fnw, ATLAS:2015hzw, ALICE:bo}, which suggest the presence of collective effects in small collision systems~\cite{ALICE:2018pal}. These observations challenge the current understanding of hadronic collisions, as different particle production mechanisms are expected to be involved in the different collision systems~\cite{PhysRevLett.48.1066}.

Several theoretical approaches have attempted to describe the strange hadron production in hadronic collisions. 
A qualitative description of the experimental results has been achieved with different event generators, such as \textsc{Pythia}8 with colour ropes~\cite{Bierlich:2014xba}, HERWIG~\cite{HERWIG, HERWIG7} and EPOS~LHC~\cite{Pierog:2013ria}, which combine perturbative Quantum Chromodynamics (pQCD) calculations with phenomenological models for the description of hard and soft processes, respectively. 
A qualitative description of strange
hadron production is also provided by the statistical hadronisation model, according to which the relative abundances of strange hadrons with respect to lighter flavours are diminished in small systems by a canonical suppression of the strangeness quantum numbers~\cite{Rafelski:2001bu, Redlich:527728, Hamieh:2000tk, Tounsi:2002nd}.
However, none of these theoretical approaches provides a consistent quantitative description of the multiplicity dependence of the hadron-to-pion ratios~\cite{ALICE:2016fzo, ALICE:2020nkc, ALICE:2022wpn}, indicating that the microscopic origin of strangeness enhancement with multiplicity in small collision systems remains an open issue.

One way to investigate this phenomenon consists in studying the strange hadron production associated with hard scattering processes and with the underlying event. 
Hard scattering processes are associated with high-energy parton shower (jet) hadronisation, whereas the underlying event consists of all the processes different from the hardest partonic interaction.
The ALICE Collaboration has recently studied the \pt~spectra of different (multi-)strange hadrons in jets and in the underlying event in minimum bias pp collisions at $\sqrt{s}=7$~TeV and \mbox{$\sqrt{s}=13$~TeV} and in p--Pb at $\sqrt{s_{\mathrm{NN}}}=5.02$~TeV~\cite{ALICE:2021vxl, ALICE:2022ecr}, using the anti-$k_{\mathrm{T}}$ algorithm~\cite{Cacciari:2008gp, Cacciari:2011ma} for jet reconstruction.
The results indicate that jet fragmentation alone is not sufficient to describe strange particle production in hadronic collisions at LHC energies and suggest that the baryon-over-meson yield ratios increase with multiplicity at intermediate \pt values~\cite{ALICE:2018pal} might be driven by particle production in the underlying event.

This paper presents a complementary measurement of strange hadron production associated with hard scattering processes and the underlying event as a function of the charged-particle multiplicity in pp collisions. 
The \pt spectra and the \pt-integrated yields of \kzero and \Xis are measured at central rapidities in \ppfi and at \thTeV in the direction of the leading particle (trigger particle), which is considered to be a proxy for the jet axis, and in the direction transverse to the trigger particle, which is associated with the underlying event and might also receive a contribution from \mbox{low-\pt} jets (\mbox{mini-jets}). 
For this purpose, the angular correlations between trigger particles and \kzero (\Xis) are exploited.
The single-strange meson \kzero and the double-strange baryon \Xis are studied as they have a different strangeness content and therefore a different sensitivity to strangeness enhancement. In addition, these species receive a negligible feed-down from other particles, which simplifies the measurement of their yields.
The \kzero and \Xis per-trigger yields per unit \dEtadPhi are reported as a function of the charged-particle multiplicity and are compared with the predictions of three different phenomenological models, namely \textsc{Pythia}8.2 with the Monash 2013 tune~\cite{Sjostrand:2014zea}, \textsc{Pythia}8.2 with ropes~\cite{Bierlich:2014xba} and EPOS LHC~\cite{Pierog:2013ria}.

The paper is organised as follows.
Section~\ref{sec:exp} outlines the experimental setup and the data sample used for this measurement, Sec.~\ref{sec:analysis} presents the experimental details of the analysis along with the associated systematic uncertainties, and Sec.~\ref{sec:results} shows the per-trigger \pt spectra and \pt-integrated yields of \kzero and \Xis as a function of the charged-particle multiplicity, together with their comparison with model predictions. Finally, the conclusions are drawn in Sec.~\ref{sec:outlook}.

\section{Experimental setup and data selection}
\label{sec:exp}
The ALICE apparatus~\cite{ALICEexp, ALICE:2014sbx} consists of central barrel detectors covering the pseudorapidity interval $|\eta| < 0.9$, a muon spectrometer covering $-4.0 < \eta < -2.5$, and a set of detectors at forward and backward rapidities used for triggering and event characterisation purposes. The central barrel detectors are positioned inside a solenoidal magnet providing a 0.5~T magnetic field along the beam axis and are used for primary vertex (PV) reconstruction, track reconstruction and charged-particle identification.
The main detectors used for 
the analysis presented in this paper 
are the Inner Tracking System (ITS)~\cite{ITStdr}, the Time Projection Chamber (TPC)~\cite{TPCtdr},
the Time Of Flight (TOF) detector~\cite{TOFtdr}, and the V0 detectors~\cite{V0Performance}.
The ITS is the innermost detector of the ALICE experiment. The ITS used during the LHC Run~2 consisted of six cylindrical layers of silicon tracking detectors placed at a radial distance from the beam pipe between 3.9 and 43.0~cm. The two innermost layers of the ITS were equipped with Silicon Pixel Detectors (SPD), the two intermediate layers consisted of Silicon Drift Detectors (SDD), and the two outermost layers of Silicon Strip Detectors (SSD). 
The SPD was used to reconstruct the PV of the collision and the tracklets, short two-point track segments covering the pseudorapidity region $|\eta|<1.2$. 
The other main functions of the ITS are the reconstruction of secondary vertices from weak decays and the tracking and identification of particles with momentum smaller than 200~MeV/\textit{c}. 
The TPC is the main tracking detector of the central barrel. It is used for the identification of charged particles by measuring the specific ionisation energy loss \dEdx. 
The TPC has a cylindrical shape with an inner radius of 85~cm, an outer radius of 250~cm and an overall length along the beam direction of~5~m. It is filled with nearly $\mathrm{90~m^3}$ of gas mixture, consisting of $\mathrm{Ar/CO_2}$ (88/12) in 2016 and 2018 and $\mathrm{Ne/CO_2/N_2}$ (90/10/5) in 2017.
It covers the pseudorapidity region of $\mathrm{|\eta| <0.9}$ for tracks with full radial length and provides full azimuthal acceptance. The TPC is radially segmented into ``pad rows":
tracks reconstructed with the TPC may consist of up to 159 points, each corresponding to one crossed pad row.
The TOF detector is an array of multigap resistive plate chambers (MRPCs) covering the pseudorapidity range of $|\eta|\lesssim0.9$ and providing full azimuthal acceptance. 
Its primary purpose is the identification of particles with intermediate momentum via the measurement of their time of flight.
The V0 detector consists of two arrays of scintillation counters, V0A and V0C, placed at forward rapidity. The V0A is located at $+$3.3~m from the interaction point and covers the pseudorapidity range of $\mathrm{2.8 < \eta <5.1}$, whereas the V0C is placed on the opposite side at $-$0.9~m from the interaction point and covers the pseudorapidity range of $\mathrm{-3.7 < \eta <-1.7}$. 
The V0 detector provides the minimum bias trigger in pp, p--Pb and Pb--Pb collisions.
It is used to classify pp collisions in multiplicity percentile classes based on the total deposited charge (V0M amplitude).

\sloppy The analysis presented in this paper was performed using pp collisions at \fiTeV and \thTeV collected by the ALICE experiment during the LHC Run~2 data-taking campaign \mbox{(2015--2018)}.
Two samples of pp collisions at \thTeV were used: one collected with the minimum bias (MB) trigger, the other collected with the high multiplicity (HM) trigger.
The MB trigger is provided by the combined signals in the V0A and V0C detectors.
The HM trigger is activated online when the amplitude of the signal in the V0 detectors is above a predefined threshold and allows for the selection of events characterised by approximately 30 charged particles produced at midrapidity, i.e., four times more than those collected in minimum bias events ($\approx$ 7).
The sample of pp collisions at \fiTeV was collected with the MB trigger, and consists of events characterised by approximately 6 charged particles produced at midrapidity.

To ensure uniform detector acceptance, the reconstructed PV position must lie within $\pm$10~cm from the nominal interaction point in the beam direction.
The contamination from in-bunch pileup events is removed by excluding events with multiple vertices reconstructed with the SPD.
The background from beam-gas events is removed by using the timing information in the V0 detectors and the correlation between SPD tracklets and SPD clusters, as discussed in detail in Ref.~\cite{ALICE:2014sbx}.

The MB events used for the trigger particle-\kzero correlation analysis were collected in 2016 and 2017 and amount to about $1\times10^9$ good quality events. As \Xis are approximately fifteen times less abundant than \kzero, all MB events collected in 2016, 2017 and 2018 were used for the trigger particle-\Xis correlation analysis, corresponding to $1.6\times10^9$ events after the quality selections. 
The sample of HM events at \thTeV\ consists of $4\times10^8$ selected events collected in 2016, 2017 and 2018.
The sample of MB pp collisions at \fiTeV\ consists of $9\times10^8$ good events recorded in 2017.

\section{Analysis details}
\label{sec:analysis}
The selected events are divided into V0M multiplicity percentile classes defined starting from the distribution of the sum of the signal amplitudes measured with the two V0 detectors.
Minimum bias events at \thTeV are divided into five multiplicity classes (0--5\%, 5--10\%, 10--30\%, 30--50\%, 50--100\%): the \mbox{0--5\%} class, for example, contains the 5\% of events with the highest V0M amplitude, while the \mbox{70--100\%} class contains the 30\% of events with the smallest V0M amplitude. 
Once corrected for the V0M trigger efficiency, these ranges represent fractional intervals of the cross section of $\mathrm{INEL>0}$ events, defined as events having at least one charged particle produced in the pseudorapidity interval $|\eta| < 1$.
The corrected intervals are respectively: 0--4.57\%, 4.57--9.15\%, 9.15--27.50\%, 27.50--46.12\%, 46.12--100\%, respectively. The details about the correction procedure can be found in Ref.~\cite{ALICE:2020swj}.
High multiplicity events at \thTeV\ are selected in the multiplicity range 0--0.1\%, which includes the 0.1\% of the MB events characterised by the highest V0M amplitude. These events are further divided into three multiplicity classes: 0--0.01\%, 0.01--0.05\% and 0.05--0.1\%, with the first one corresponding to 0--0.0091\% of the $\mathrm{INEL>0}$ cross section, and the sum of the other two classes to 0.0091--0.0915\%.
The available number of MB events at \five allows for the analysis to be performed only in two multiplicity classes (0--10\%, 10--100\%, corresponding to 0--9.15\% and 9.15--100\% of the $\mathrm{INEL>0}$ cross section, respectively), as at this energy the sample of MB events is smaller and the average strange hadron yields per event are smaller than those at \thTeV. 
For each V0M percentile class, the average multiplicity of charged particles produced at midrapidity in events containing a trigger particle,  \dNdetatrigg, and its systematic uncertainties are computed using the technique described in Ref.~\cite{ALICE:2020swj}. 

\subsection{Trigger particle identification}
\label{sec:trigger}
In this analysis, a trigger particle is defined as the charged particle with the highest-\pt in a given event (leading particle), coming from the PV, produced in the pseudorapidity interval $|\eta| < 0.8$ and within the transverse momentum range \mbox{$3 < p_{\mathrm{T}} < 15$~GeV/\textit{c}}. The minimum \pt threshold is applied to select particles originating from the hadronisation of hard scattering processes. 
An increase of the threshold value above 3~GeV/\textit{c} would increase the contribution from particles originating from hard scattering processes. However, it would also decrease the number of events with a trigger particle, limiting the possibility of performing a multiplicity dependent measurement of the angular correlation between trigger particles and \Xis baryons.
The trigger particles are selected starting from the tracks reconstructed using the TPC and constrained to the PV.
Only tracks in the $\mathrm{|\eta|< 0.8}$ acceptance  region, where full track reconstruction is provided, are accepted.
Standard selections are applied: tracks are required to cross at least 80 out of 159 TPC pad rows and to be formed by more than 70 TPC clusters, where a cluster is the signal induced by the passage of the particle in a crossed pad row. In order not to have large gaps in the number of expected tracking points in the radial direction, the ratio of crossed pad rows $N_\mathrm{crossed}$ over findable clusters $N_\mathrm{findable}$ is required to be greater than 0.8.
In order to reject the low-resolution tracks which pass through the edges of the TPC sectors, tracks with radial lengths smaller than 90~cm are discarded,
and the ratio between the number of crossed pad rows and the radial track length is required to be greater than 0.8~$\mathrm{cm}^{-1}$. 
In addition, a maximum \pt threshold of 15~GeV/\textit{c} is applied to retain only tracks with \pt resolution better than 2\%. This selection rejects less than 0.5\% of tracks.
Finally, the goodness-of-fit $\chi^2$ per TPC cluster of the track fit in the TPC is required to be smaller than~4.
To discard charged particles not originating from the PV, a selection on the distance of closest approach (DCA) of the track to the PV is applied both along the beam direction $z$ ($\mathrm{DCA}_z$) and in the perpendicular plane ($\mathrm{DCA}_{xy}$):
\begin{equation*}
\label{eq:dcacut}
        \mathrm{|DCA_\emph{z}| < 0.04~cm \;, \> |DCA_\emph{xy}| < \left(0.0105 + \frac{0.035}{[\textit{p}_T/(GeV/\textit{c})]^{1.1}}\right)~cm}.
\end{equation*}
The \pt-dependent selection on the $\mathrm{DCA}_{xy}$ allows for selecting tracks within 7$\sigma$ from the interaction vertex in the transverse plane, where $\sigma$ is the resolution with which the $\mathrm{DCA}_{xy}$ is measured.\\
The fraction of good-quality events containing a trigger particle increases with the event multiplicity, from approximately 2\% in the 50--100\% V0M class to 50\% in the highest multiplicity class 0--0.01\%, as it is more likely to find a high-\pt track
in events characterised by a larger multiplicity of charged particles.

\subsection{Identification of \kzero and \Xis}
The strange hadrons \kzero, \Ximinus and \Xiplus (in the following \Xis) are identified in the pseudorapidity range of $|\eta|<0.8$ via invariant mass analysis techniques, exploiting the topology of their weak decays into charged hadrons~\cite{Workman:2022ynf}:

\begin{minipage}{0.5\textwidth}
\begin{flalign*}
\qquad
& \mathrm{K^0_S} \rightarrow \pi^+ \pi^- &\\
& \Xi^- \rightarrow \Lambda \pi^- \rightarrow \mathrm{p} \pi^- \pi^- 
    (\Xi^+ \rightarrow \overline{\Lambda} \pi^+ \rightarrow \mathrm{\overline{p}} \pi^+ \pi^+) &\\
& &\\
\end{flalign*}
\end{minipage}
\begin{minipage}{0.5\textwidth}
\begin{flalign*}
\qquad
& \mathrm{B.R. = (69.20 \pm 0.05)} \% &\\
& \mathrm{B.R. (\Xi^- \rightarrow \Lambda \pi^-) = (99.887 \pm 0.035)} \% &\\
& \mathrm{B.R. (\Lambda \rightarrow p \pi^-) = (63.9 \pm 0.5)} \%. 
\end{flalign*}
\end{minipage}
The charged daughter tracks of \kzero and \Xis candidates are selected in the pseudorapidity range of \mbox{$|\eta| < 0.8$}, and are required to satisfy the same track quality criteria applied for the trigger particle selection. 
Daughter tracks in the whole \pt interval are identified by requiring the specific ionisation energy loss $\mathrm{d}E/\mathrm{d}x$ measured in the TPC to be compatible with the expected theoretical value within $\pm3\sigma$, where $\sigma$ is related to the resolution with which $\mathrm{d}E/\mathrm{d}x$ is measured.
In addition, daughter tracks are required not to be associated with a ``kink topology"~\cite{ALICE:2011gmo}, which is characteristic of the decay of charged kaons.
The combinatorial background is suppressed by applying standard topological selections (see ref.~\cite{ALICE:2019avo}), listed in Table~\ref{tab:decayKsXi}.
\begin{table}[!htbp]
\caption{Daughter-track quality selections, topological and kinematic selections applied to \kzero and \Xis candidates.
The symbol $\theta_\mathrm{P}$ stands for the pointing angle, i.e. the angle between the reconstructed momentum vector of the \kzero and \Xis candidates and the line connecting the primary to the secondary vertex. 
All other symbols are explained in the text.
}
\begin{center}
\renewcommand{\arraystretch}{1.15} 
\begin{tabular}{|l|c|}
\hline
\multicolumn{2}{|c|}{\textbf{Daughter-track selections}}\\
\hline
\hline
{Number of TPC clusters} & {$>$ 70} \\
\hline
{$\chi^2$/ndf} & {$<$ 4} \\
\hline
{Number of TPC crossed pad rows $N_\mathrm{crossed}$} & {$>$ 80} \\
\hline
{$N_\mathrm{crossed}/N_\mathrm{findable}$} & {$>$ 0.8} \\
\hline
{Track length $l_\mathrm{TPC}$ in the TPC} & {$>$ 90~cm} \\
\hline
{$N_\mathrm{crossed}/l_\mathrm{TPC}$}& {$>$ 0.8 $\mathrm{cm}^{-1}$}  \\
\hline
{Rejection of kink topology} & {Yes} \\
\hline
{$|\eta|$} & {$<$ 0.8} \\
\hline
{$\mathrm{d}E/\mathrm{d}x$ measured in the TPC} & {$<3\sigma$} \\
\hline
{At least one daughter track has a hit in the SPD or in the TOF} & {Yes} \\
\hline
\hline
\multicolumn{2}{|c|}{$\mathbf{K^0_S}$-\textbf{topological-variable selections}} \\
\hline
\hline
{DCA daughter tracks to PV} & {$>$ 0.06 cm} \\
\hline
{DCA between daughter tracks} & {$<$ 1$\sigma$} \\
\hline
{$\mathrm{cos(\theta_P)}$} & {$>$ 0.995} \\
\hline
{DCA \kzero to PV} & {$<$ 0.5 cm} \\
\hline
{\kzero decay radius} & {$>$ 0.5 cm}\\
\hline
\hline
\multicolumn{2}{|c|}{\textbf{\Xis-topological-variable selections}} \\
\hline
\hline
{DCA meson daughter to PV} & {$>$ 0.04 cm} \\
\hline
{DCA baryon daughter to PV} & {$>$ 0.03 cm} \\
\hline
{DCA bachelor to PV} & {$>$ 0.04 cm} \\
\hline
{DCA between daughter tracks of the $\Lambda$} & {$<$ 1.5$\sigma$} \\
\hline
{$\mathrm{cos(\theta_P)}$ (of \Xis to PV)} & {$>$ 0.995} \\
\hline
{$\mathrm{cos(\theta_P)}$ (of (anti-)$\Lambda$ to \Xis decay vertex)} & {$>$ 0.97} \\
\hline
{DCA between bachelor and (anti-)$\Lambda$} & {$<$ 0.8 cm} \\
\hline
{DCA $\Lambda$ to PV} & {$>$ 0.06 cm} \\
\hline
{(anti-)$\Lambda$ decay radius} & {$>$ 1.1 cm}\\
\hline
{\Xis decay radius} & {$>$ 0.5 cm}\\
\hline
\hline
\multicolumn{2}{|c|}{$\mathbf{K^0_S}$-\textbf{candidate selections}} \\
\hline
\hline
{$|\eta_{\rm K^0_S}|$} & {$<$ 0.8}\\
\hline
{$\mathrm{|m_{\pi p} - m_{\Lambda}|}$} & {$>5$~MeV/$c^2$} \\
\hline
{Proper lifetime $\tau$} & {$<20$~cm/\textit{c} ($\simeq 7.5\langle \tau_{\kzero}\rangle$)} \\
\hline
\hline
\multicolumn{2}{|c|}{\textbf{\Xis-candidate selections}}   \\
\hline
\hline
{$|\eta_{\Xi^\pm}|$} & {$<$ 0.8}\\
\hline
{$\mathrm{|m_{\pi p} - m_{\Lambda}|}$} & {$<6$~MeV/$c^2$} \\
\hline
{$\mathrm{|m_{K \Lambda} - m_{\Omega}|}$} & {$>5$~MeV/$c^2$} \\
\hline
{Proper lifetime $\tau$} & {$<14.73$ cm/\textit{c} (= 3$\langle \tau_{\Xis}\rangle$)} \\
\hline
\end{tabular}
\medskip
\label{tab:decayKsXi}
\end{center}
\end{table}
A selection on the proper lifetime $\tau$ of \kzero and \Xis candidates is also applied. 
The proper lifetime is calculated as $\tau = d \times m /|\vec{p}|$, where $m$ is the nominal mass of the considered particle, $|\vec{p}|$ is the magnitude of the reconstructed momentum, and $d$ is the distance of the reconstructed secondary decay vertex from the primary one. 
In order to identify \Xis, the invariant mass of the daughter (anti-)$\Lambda$ is required to differ from the nominal mass value of the $\Lambda$ by less than 
6~MeV/$c^2$, according to the (anti-)$\Lambda$ invariant mass resolution.
The background from (anti-)$\Lambda$ in the \kzero sample is suppressed by rejecting the \kzero candidates whose invariant mass calculated under the p$\pi$ assumption for the daughter tracks lies within $\pm5$~MeV/$c^2$ from the nominal $\Lambda$ mass.
Similarly, the background from $\Omega^{\pm}$ in the cascade sample is tackled by rejecting the cascade candidates whose invariant mass calculated under the $\Lambda \mathrm{K}$ assumption for the daughter particles lies within $\pm5$~MeV/$c^2$ from the nominal $\Omega$ mass.
The width of the rejected region is determined according to the invariant mass resolution $\sigma$ of the competing candidate, and corresponds to approximately $\pm3\sigma$.
Finally, to reduce the out-of-bunch pileup background caused by tracks from other bunch crossings within the TPC integration time, at least one of the daughter tracks is required to have a hit in the TOF or the SPD. 

The signal extraction is performed as a function of \pt. 
The invariant mass distributions of \kzero and \Xis candidates are fitted with the sum of two Gaussian functions, used to take into account the invariant mass resolution of the signal peak, and a first-degree polynomial, used to describe the background. 
A ``peak" region is defined within $\pm4\sigma$ from $\mu$, where $\mu$ and $\sigma$ are the average mean value and width of the two Gaussian functions, respectively. 
For each candidate, ``sideband" regions are defined:
the sidebands of the \kzero (\Xis) invariant mass distributions are defined as the intervals \mbox{$\mu - 10\sigma <m_{\pi^+ \pi^-} (m_{\pi \Lambda}) < \mu - 4\sigma$} and \mbox{$\mu + 4\sigma <m_{\pi^+ \pi^-} (m_{\pi \Lambda})< \mu + 10\sigma$}.
The purity of the \kzero and \Xis candidates samples, defined as the ratio between the signal and the total number of candidates in the invariant mass range within $\pm4\sigma$ from $\mu$, is larger than 0.95 and 0.89 for \kzero and \Xis, respectively.

\subsection{The angular correlation function}
\label{subsec:correlation}
\begin{figure}[htbp!]
\centering
{\includegraphics[scale=0.8]
{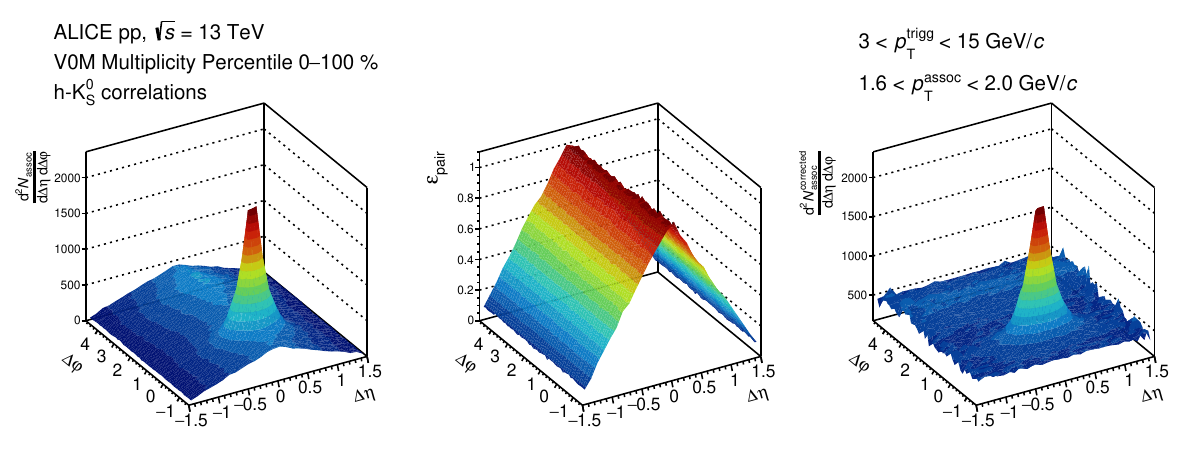}}
\caption{(left) Example of angular correlation distribution between trigger and \kzero found in the same collision. (centre) Acceptance correction of trigger-\kzero pairs. (right) Angular correlation distribution divided by the pair acceptance. 
}
\label{fig:AC}   
\end{figure}

\begin{figure}[htbp!]
\centering
{\includegraphics[scale=0.8]{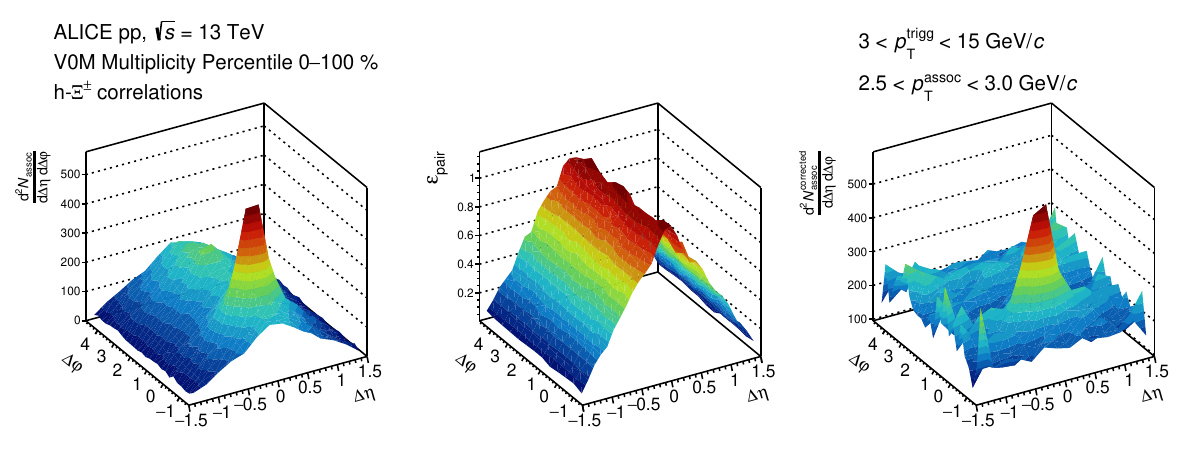}}
\caption{(left) Example of angular correlation distribution between trigger and \Xis found in the same collision. (centre) Acceptance correction of trigger-\Xis pairs. (right) Angular correlation distribution divided by the pair acceptance. 
}
\label{fig:ACXi}   
\end{figure}

The angular correlation between trigger particles, denoted as ``h", and associated particles, i.e. the
\kzero (\Xis) candidates with an invariant mass within 4$\sigma$ from the average mean value $\mu$ of the Gaussian fit functions, is expressed as a function of the pseudorapidity difference \dEta and the azimuthal angle difference \dPhi between the trigger and associated particles.
Examples of the angular correlation distribution $\mathrm{d}^2{N_{\mathrm{assoc}}}(\Delta\eta, \Delta\varphi)/\mathrm{d}\Delta\eta \mathrm{d}\Delta\varphi$
of \hKs and \hXi pairs produced in \ppmb are shown in the left panel of Figs.~\ref{fig:AC} and~\ref{fig:ACXi}, respectively. 
The distributions show a near-side peak centred at ($\Delta\eta$, $\Delta\varphi$) = (0, 0) which is associated with h-\kzero and h-\Xis pairs fragmented within the same jet.
The distributions are corrected by the efficiency$\times$acceptance$\times$B.R. of associated particles $\epsilon_\mathrm{assoc}$ 
computed using a Monte Carlo simulation based on \textsc{Pythia}8.2 with the Monash 2013 tune~\cite{Sjostrand:2014zea} for the generation of events and on \textsc{Geant}~4~\cite{Agostinelli:2002hh} for the description of the propagation of particles through the material of the detector.
The term $\epsilon_\mathrm{assoc}$ is calculated in events with a trigger particle identified by applying the selections described in Sec.~\ref{sec:trigger}. 
It increases with \pt, reaching a saturation value of about 35\% and 25\% at $p_\mathrm{T} = 3$ and 4~GeV/\textit{c} for \kzero and \Xis, respectively. 
For \kzero, $\epsilon_\mathrm{assoc}$ increases with decreasing charged-particle multiplicity, varying by about 10\% from the 0--5\% to the 50--100\% V0M multiplicity classes, whereas for \Xis, because of the different decay topology, it does not depend on multiplicity.
For both particles, $\epsilon_\mathrm{assoc}$ is computed in each multiplicity class and as a function of $\eta$ and \pt, and is applied as a weight factor to each entry of the angular correlation distribution.

The angular correlation distributions (left panel of Figs.~\ref{fig:AC} and~\ref{fig:ACXi}) exhibit a triangular shape in \dEta, which is related to the geometrical acceptance of the trigger--associated particle pairs. It is corrected for by the pair acceptance $\epsilon_\mathrm{pair}$, calculated with the mixed-event method, which correlates the trigger particle found in one event with the associated particles produced in different events. 
These events are required to have similar characteristics, namely to lie in the same multiplicity class, to have the z-coordinate of the PV differing by less than 2~cm, and to contain a trigger particle. 
Each entry of the mixed-event angular correlation distribution is weighed with $1/\epsilon_\mathrm{assoc}$, to take into account the $\eta$ dependence of the associated particle efficiency.
As shown in the central plot of~Figs.~\ref{fig:AC} and~\ref{fig:ACXi}, the mixed-event angular correlation distribution has a triangular shape in $\Delta \eta$, determined by the $\eta$ acceptance. In contrast, it shows no dependence on \dPhi, as a consequence of the cylindrical symmetry of the detector. 
To obtain the pair acceptance, the mixed-event distribution is normalised to unity at $\Delta \eta \simeq 0$, where all particle pairs are assumed to be accepted. The raw angular correlation distributions are divided by the pair acceptance to retain the genuine physical correlations in such pair-acceptance window $\mathrm{d}^2{N^\mathrm{corrected}_{\mathrm{assoc}}}(\Delta\eta, \Delta\varphi)/\mathrm{d}\Delta\eta \mathrm{d}\Delta\varphi$, shown in the right panel of Figs.~\ref{fig:AC} and~\ref{fig:ACXi}. The pair acceptance is computed in each multiplicity class. Since the \hXi acceptance does not show any multiplicity dependence within the statistical uncertainty, the correction is performed using the pair acceptance computed in the 0-100\% multiplicity class, in order to reduce statistical fluctuations.

\subsection{Evaluation of the \texorpdfstring{\pt}{pT} spectra and integrated yields}
The corrected angular correlation distributions $\mathrm{d}^2{N^\mathrm{corrected}_{\mathrm{assoc}}}(\Delta\eta, \Delta\varphi)/\mathrm{d}\Delta\eta \mathrm{d}\Delta\varphi$ (right panel of Figs.~\ref{fig:AC} and~\ref{fig:ACXi}) are projected onto the \dPhi axis.
The \dPhi projections are corrected for the contribution of the combinatorial background due to candidates which are not \kzero (\Xis).
The standard procedure, which is applied to \kzero in all samples and to \Xis in the HM sample, consists of subtracting the angular correlations obtained using \kzero (\Xis) candidates in the ``sidebands" of the invariant mass distributions from those obtained from the peak region. Before subtraction, the angular correlation obtained from the sidebands is divided by the integral of the invariant mass distribution in the sidebands regions and multiplied by the integral of the background fit function in the signal region in order to take into account the purity of the \kzero and \Xis samples.
A different procedure is applied to take into account the contribution of misidentified \Xis in minimum bias \ppmb
and at \fiTeV, since these samples do not contain enough events to compute the \hXi angular correlation distributions from the sidebands regions. In this case, the \dPhi projections are multiplied by the purity of the sample of \Xis candidates. This procedure assumes that the angular correlation distributions for background candidates have the same shape as for signal candidates. A systematic uncertainty is applied to take into account any difference with respect to the standard procedure, as described in Sec.~\ref{sec:sys}.

In addition, the \dPhi projections are corrected for the fraction of feed-down \kzero (\Xis). 
For this purpose, the distributions are multiplied by ($1-F_\mathrm{NP}$), where $F_\mathrm{NP}$ is the fraction of non-primary \kzero (\Xis) calculated using Monte Carlo simulations. This procedure is based on the assumption that the angular correlation for feed-down particles has the same shape as the angular correlation for primary strange hadrons. This correction has a negligible impact since in the \pt ranges considered in this analysis $F_\mathrm{NP}\sim0.5\%$ for \Xis and $F_\mathrm{NP}<0.05\%$ for \kzero.

The associated particle yields are computed by integrating the \dPhi projections and are divided by the width of the \dEtadPhi region from which they are extracted. 
The toward-leading production is extracted from the region ($|\Delta \eta| < 0.86$, $|\Delta \varphi| < 1.1$), chosen to include the whole near-side peak. 
The \mbox{transverse-to-leading} production is extracted from \mbox{($0.86 < |\Delta \eta| < 1.2$, $0.96 < \Delta \varphi < 1.8$)}: this region is chosen to exclude the away-side peak associated with the recoil jet, whose contribution is situated around $\Delta\varphi \sim \pi$ and is elongated over the whole \dEta interval, and to exclude any possible residual near-side peak contribution in the region around $\Delta\varphi\sim0$. 
Finally, the full yield is obtained from the whole \mbox{$\Delta\eta\Delta\varphi$} region \mbox{($|\Delta \eta| < 1.2$, $-\pi/2 < \Delta \varphi < 3/2\pi$)}.

To obtain the toward-leading yield, the contribution of the underlying event is subtracted from the \tl \dPhi projections.
An estimate of the underlying event contribution is provided by
the long-range  \dPhi projections obtained from the $0.86<|\Delta \eta|<1.2$ region
and scaled to take into account the different \dEta widths of the two regions.
This procedure cannot be applied to extract the \tl yield of \Xis with \mbox{\pt $\lesssim$ 2 GeV/\textit{c}} in the minimum bias samples because of the large statistical uncertainties affecting the long-range \dPhi projections.
To overcome this issue, the angular correlation between charged particles with $0.15 < p_\mathrm{T} < 2.5$~GeV/\textit{c} and \Xis candidates is computed, and the \dPhi projections obtained from the $0.86<|\Delta \eta|<1.2$ region are used as estimates of the underlying event, after being scaled in order to match the $|\Delta\eta|<0.86$ projections in the interval $1 \lesssim \Delta\varphi \lesssim 2$.
These projections do not suffer from large statistical uncertainties. They are observed to be compatible within uncertainties with the default distributions in the $-\pi/2 < \Delta\varphi < \pi/2$ interval, where the near-side peak lies.
A systematic uncertainty related to this procedure is evaluated as described in Sec.~\ref{sec:sys}.

The per-trigger yields per unit \dEtadPhi, from now on referred to as ``yields", are corrected by an additional normalisation factor $C_\mathrm{norm}$ in order to obtain the fully corrected \pt spectra $\left(\frac{1}{N_{\mathrm{trigg}}} \frac{1}{\Delta\eta\Delta\varphi} \frac{\mathrm{d}N}{\mathrm{d}p_{\mathrm{T}}}\right)$ in the three different regions:

\begin{equation}
\label{eq:ptspectra}
\frac{1}{N_{\mathrm{trigg}}} \frac{1}{\Delta\eta\Delta\varphi} \frac{\mathrm{d}N}{\mathrm{d}p_{\mathrm{T}}}\ = \frac{1}{N_{\mathrm{trigg}}} 
\frac{1}{\Delta\eta\Delta\varphi} \frac{1}{\Delta p_{\mathrm{T}}}
C_\mathrm{norm}\int_{\Delta\varphi} \frac{\mathrm{d}N^\mathrm{corrected}_{\mathrm{assoc}}}{\mathrm{d\Delta\varphi}} \mathrm{d\Delta\varphi},
\end{equation}

where $N_\mathrm{trigg}$ is the number of trigger particles in a given V0M multiplicity class and $C_\mathrm{norm}$ considers the efficiency with which events with a trigger particle are selected. 
The normalisation factor $C_\mathrm{norm}$ is computed using a Monte Carlo simulation and depends on the efficiency of trigger particle reconstruction and the difference between the \kzero (\Xis) spectra measured in events with a reconstructed trigger particle and events with a generated trigger particle. This correction factor is compatible with one for the \tl spectra, whereas it decreases with \pt for full and \tr spectra, reaching a saturation value of about 0.98 at $p_\mathrm{T} > 3$~GeV/\textit{c} for both \kzero and \Xis.

To compute the \pt-integrated yields, the spectra are fitted with four different functions used to extrapolate the yield in the unmeasured \pt interval.
The extrapolated yield is the average obtained from the four different fit functions: the L\'evi-Tsallis~\cite{Tsallis:1987eu}, the Boltzmann, the Fermi-Dirac, and the $m_{\mathrm{T}}$-exponential functions~\cite{ALICE:2019avo}.
The extrapolated fraction of the \kzero yield amounts up to approximately 1\% of the total yield for full and \tr production and to approximately 8\% for \tl production, because of the larger unmeasured \pt interval.
The extrapolated fraction of the \Xis yield varies between 10\% and 40\% for full and \tr production and between 20\% and 35\% for \tl production, depending on the multiplicity class. It is worth mentioning that the extrapolated fraction for transverse-to-leading yields is larger than for toward-leading yields in the same unmeasured \pt interval, as transverse-to-leading spectra are softer.

\subsection{Systematic uncertainties}
\label{sec:sys}
Several systematic uncertainties affecting the measurement of the full, \tr and \tl \pt spectra are investigated.
All the considered sources of systematic uncertainties are reported in Table~\ref{tab:sysPtSpectra} for \kzero (top) and \Xis (bottom) \pt spectra, together with the relative uncertainty associated with each of the sources at three different \pt values in minimum bias \ppmb. 

\begin{table}[!htbp]
\caption{Summary of the relative systematic uncertainties of the \kzero (top) and \Xis (bottom) \pt spectra measured in \ppmb in the V0M multiplicity class 0-100\%. The values in parentheses refer to the \tl spectra and are reported only when a difference from the \tr and full spectra is observed. 
No systematic uncertainty for the \tl \Xis spectra is reported in the lowest \pt interval, as the measurement is performed for $p_{\mathrm{T}}>1.0(1.5)$~GeV/\textit{c}, depending on the multiplicity class. No significant \mbox{centre-of-mass} energy dependence is observed. The three sources of uncertainty marked with an asterisk are observed to be partially uncorrelated across multiplicity, whereas all the other sources are fully correlated across multiplicity.
See text for details.}
\begin{center}
\renewcommand{\arraystretch}{1.35} 
\resizebox{\textwidth}{!}{
\begin{tabular}{|p{0.4\textwidth}|>{\centering}p{0.2\textwidth}>{\centering}p{0.2\textwidth}>{\centering\arraybackslash}p{0.2\textwidth}|}
\hline
\multicolumn{1}{|c|}{{\textbf{Hadron}}} &
\multicolumn{3}{c|}{{$\mathrm{K^0_S}$}}\\
\multicolumn{1}{|c|}{$p_\mathrm{T}$ (GeV/\textit{c})} &
$\approx$0.6 & $\approx$1.8 & $\approx$3.5 \\
\hline
\hline
{Topological selections:*} & {} & {} & {}\\
\hline
{\textit{Full}} & {0.3\%} & {0.3\%} & {0.3\%}\\
\hline
{\textit{Transverse-to-leading}} & {0.5\%} & {0.5\% } & {0.5\%}\\
\hline
{\textit{Toward-leading}} & {2\%} & {2\%} & {1\%}\\
\hline
{Trigger particle $\mathrm{DCA}_z$ selection} & {0.1\%} & {0.07\%} & {0.05\%}\\
\hline
{Choice of \dEta region*} & {0.3\% (2\%)} & {0.5\% (1.2\%)} & {0.7\% (0.7\%)} \\
\hline
{Choice of \dPhi region*} & {0.7\% (2.5\%)} & {0.7\% (0.7\%)} & {1.2\% (0.2\%)} \\
\hline
{Background fit function} & {0.1\%} & {0.3\%} & {0.5\%} \\
\hline
{Choice of Monte Carlo} & {1\%} & {1\%} & {1\%}\\
\hline
{Material budget} & {2\%} & {0.2\%} & {0.4\%} \\
\hline
{Residual in-bunch pileup} & {2\%} & {2\%} & {2\%}\\
\hline
{Out-of-bunch pileup track rejection} & {1.2\%} & {1.2\%} & {1.2\%} \\
\hline
{\textbf{Total}} & \textbf{{3\% (5\%)}} & \textbf{{3\% (3.5\%)}} & \textbf{{2.5\% (3\%)}} \\
\hline
\hline
\multicolumn{1}{|c|}{{\textbf{Hadron}}} &
\multicolumn{3}{c|}{{$\Xi^\pm$}}\\
\multicolumn{1}{|c|}{$p_\mathrm{T}$ (GeV/\textit{c})} &
$\approx$0.6 & $\approx$1.8 & $\approx$3.5 \\
\hline
{Topological selections:*} & {} & {} & {}\\
\hline
{\textit{Full}} & {1\%} & {0.1\%} & {0.2\%}\\
\hline
{\textit{Transverse-to-leading}} & {3.0\%} & {0.6\% } & {0.5\%}\\
\hline
{\textit{Toward-leading}} & {--} & {5\%} & {3\%}\\
\hline
{Trigger particle $\mathrm{DCA}_z$ selection} & {0.1\%} & {0.07\%} & {0.05\%}\\
\hline
{Choice of \dEta region*} & {2\%} & {1\% (2\%)} & {1\% (1\%)} \\
\hline
{Choice of \dPhi region*} & {0.9\%} & {1\% (--)} & {1.2\% (--)} \\
\hline
{Background fit function} & {0.5\%} & {0.5\%} & {0.5\%} \\
\hline
{Misidentified \Xis\ subtraction} & {0.8\%} & {0.4\% (2.5\%)} & {0.3\% (1.2\%)}  \\
\hline
{Out--of--jet subtraction} & {--} & {5\%} & {--} \\
\hline
{Material budget} & {2\%} & {2\%} & {2\%} \\
\hline
{Residual in-bunch pileup} & {2\%} & {2\%} & {2\%} \\
\hline
{Out-of-bunch pileup track rejection} & {2\%} & {2\%} & {2\%} \\
\hline
{\textbf{Total}} & \textbf{{5\%}} & \textbf{{3\% (8\%)}} & \textbf{{3\% (4\%)}} \\
\hline
\end{tabular}
}
\label{tab:sysPtSpectra}
\end{center}
\end{table}

The topological selections are varied to take into account the differences between the distributions of the topological variables in the data and in the Monte Carlo simulation used to compute the \kzero and \Xis acceptance$\times$efficiencies. 
The systematic uncertainty is evaluated from the distribution of the fully corrected yields obtained by randomly changing the topological selections within ranges leading to a maximum variation of about $\pm2$\% in the raw signal yields when one single topological variable is varied.
The relative systematic uncertainty depends on the multiplicity class. Overall, it is smaller than 2\% (4\%) for the \tr and full \pt spectra of \kzero (\Xis). 
For the toward-leading spectra of \kzero (\Xis) it reaches values up to 8\% at $p_\mathrm{T}<1$ (2) GeV/\textit{c}, decreasing with increasing \pt. This source of uncertainty represents the dominant one for the \tl spectra. 

 The effect of a different fraction of non-primary charged particles in the sample of trigger particles is evaluated by varying the selection applied to the $\mathrm{DCA}_z$ of the trigger particles. The systematic uncertainty is extracted from the distribution of the fully corrected yields obtained by randomly changing the $\mathrm{DCA}_z$ selection within the $(0, 2)$~cm range. 
The relative uncertainty associated with this source is smaller than 0.2\% for full and \tr production, and smaller than 0.5\% for toward-leading production: it represents the smallest contribution to the total systematic uncertainty.

The systematic uncertainty associated with the choice of the \dEta region is assessed by changing the default boundaries of the \dEta regions by about $+$10\%. The boundaries are not decreased below the default value, in order not to exclude any part of the near-side peak.
The results are compared with those obtained with the default ranges. The variations are significant according to the Barlow criterion~\cite{Barlow:2002yb}, with a $2\sigma$ threshold in at least four out of ten \dPhi intervals, indicating that the probability that they are due to statistical fluctuations is smaller than 0.1\%.
For both \kzero and \Xis, the relative systematic uncertainty of the \tr spectra is smaller than 2\%, whereas for the \tl spectra it decreases with \pt from at most 6\% 
for $p_\mathrm{T} < 1 (2)$~GeV/\textit{c} for \kzero (\Xis)
to less than 2\% for $p_\mathrm{T} > 3$~GeV/\textit{c}. 
The full yield, which by definition is obtained from the region ($|\Delta \eta| < 1.2$, $-\pi/2 < \Delta \varphi < 3/2\pi$), is not affected by this source of systematic uncertainty.

Similarly, the systematic uncertainty related to the choice of the \dPhi interval is assessed by changing the default boundaries of the \dPhi regions by about $\pm10$\%. 
For both \kzero and \Xis, the variations of the \tr \dPhi interval are significant according to the Barlow check  with a $2\sigma$ threshold in at least three \pt intervals. The relative uncertainty, computed taking into account only the significant variations, increases with \pt up to 2\% for both particles.
The variations of the \tl yields are significant for \kzero in minimum bias \ppmb: the relative uncertainty decreases with \pt from about 2\% down to $\sim0.1\%$ for $p_{\mathrm{T}} > 4$~GeV/\textit{c}. On the contrary, the variations are not significant for \Xis \tl spectra.
As for the choice of the \dEta region, this source does not affect the full yields.

The relative uncertainties associated with the topological selections and the choice of the \dEta and \dPhi intervals mildly depend on the multiplicity class.

Another systematic effect is related to the choice of the function used to fit the background of the invariant mass distributions of \kzero and \Xis candidates. 
To quantify it, the fit to the background is performed with a second-degree polynomial and the invariant mass interval in which the fit is performed is varied. The results obtained in this way are compared with the default ones. For \kzero, the relative systematic uncertainty increases with \pt up to 1.5\%. For \Xis, the relative systematic uncertainty equals 0.5\% in all \pt intervals. No dependence on the multiplicity class is observed.

To account for the simplified procedure applied to the subtraction of the contribution of misidentified \Xis in the minimum bias samples, the \Xis spectra measured in the 0--100\% multiplicity class of \ppmb are compared with those obtained using the method based on the sidebands of the invariant mass distribution. 
The difference between the spectra obtained with the two methods is significant according to the Barlow criterion with a $2\sigma$ threshold in at least three \pt intervals, and their relative half-difference is assigned as a systematic uncertainty to the \Xis spectra in all multiplicity classes in minimum bias events. The relative uncertainty decreases with increasing \pt, it is smaller than 1\% for full and \tr production and smaller than 3\% for \tl production. 

Since the \kzero efficiency depends on the multiplicity, a systematic uncertainty is assigned to \kzero spectra in order to account for possible differences between the multiplicity distribution in the data and in the Monte Carlo simulation used to compute the efficiency correction.
To assess this uncertainty, the default \kzero efficiencies, computed using a Monte Carlo distribution based on \textsc{Pythia}8, are compared with those obtained using a different Monte Carlo simulation based on EPOS~LHC~\cite{Pierog:2013ria}, and a 1\% uncertainty is added to account for the differences.

Another source of uncertainty for the \Xis \tl spectra is related to the method applied to subtract the contribution of the underlying event in the low-\pt intervals ($p_\mathrm{T} < 2.5$~GeV/\textit{c}) where the standard method cannot be applied due to large statistical uncertainties. To evaluate this uncertainty, the \Xis \tl spectra are compared with those obtained using the standard procedure in the \mbox{[2.0-2.5)}~GeV/\textit{c} interval, where the number of \Xis candidates is large enough to allow for the application of both methods.
The systematic uncertainty, which amounts to 5--10\% depending on the multiplicity class, is also assigned to the lower \pt intervals where the extraction procedure of the toward-leading yield differs from the standard one.

To take into account the imperfect reproduction of the detector material budget in the Monte Carlo simulation, the \kzero and \Xis efficiencies are compared with those obtained using a Monte Carlo with a different dependence of the material budget on the radial distance from the interaction point.
For \kzero, the uncertainty associated with the material budget decreases with \pt from a maximum of 2\% and shows a similar trend in all multiplicity classes. 
For \Xis, this systematic uncertainty amounts to 2\% and is independent of multiplicity and \pt.

The systematic uncertainties related to pileup rejection are inherited from the analysis of \mbox{(multi-)strange} hadron production in \ppmb~\cite{ALICE:2019avo}.
To account for a residual contamination from in-bunch pileup, a relative systematic uncertainty of 2\% is assigned to both \kzero and \Xis \pt spectra. The systematic uncertainty due to out-of-bunch pileup, evaluated in Ref.~\cite{ALICE:2019avo} by changing the matching scheme of the decay tracks with the ITS and TOF detectors, amounts to 1.2\% (2\%) for \kzero (\Xis) spectra in all \pt intervals and multiplicity classes.

Finally, another source of systematic uncertainty affecting the \pt-integrated yield is associated with choosing the fit function used to extrapolate the \pt-spectra. The uncertainty is given by the half-difference between the maximum and the minimum extrapolated yields obtained with the four different fit functions. This uncertainty amounts at most to 0.5\% (4\%) for full and \tr yields of \kzero(\Xis), and to 2\% (4\%) for \tl yields of \kzero (\Xis).   

Most of the sources of systematic uncertainties considered in this analysis are fully correlated across multiplicity, as they determine a shift of the yields in the same direction in all multiplicity classes.
Three sources of uncertainty, namely the selections applied to identify \kzero and \Xis candidates and the choices of the \dPhi and \dEta intervals, are observed to be partially uncorrelated across multiplicity. 
For each of these sources, in order to determine the fraction of uncertainty which is uncorrelated across multiplicity, the ratio $R_\mathrm{var}^{m}$ is computed:

\begin{equation}
    R_\mathrm{var}^{m} = \frac{y_\mathrm{var}^{m} / y_\mathrm{def}^{m}} {y_\mathrm{var}^{0-100\%} / y_\mathrm{def}^{0-100\%}}.
\end{equation}
Here $y_\mathrm{def}^{m}$ and $y_\mathrm{def}^{0-100\%}$ are the default yields measured in a given \pt interval in the multiplicity class~$m$ and 0--100\%, respectively, and $y_\mathrm{var}^{m}$ and $y_\mathrm{var}^{0-100\%}$ are the yields obtained by applying a systematic variation. 
If a source of uncertainty is fully correlated across multiplicity, $R_\mathrm{var}^{m}\sim1$. For each source of systematic uncertainty, the uncorrelated relative uncertainty across multiplicity is computed as the maximum deviation of $R_\mathrm{var}^{m}$ from unity.

On average, the uncorrelated fraction of the total systematic uncertainty for \kzero (\Xis) amounts to approximately 3\%(5\%), 10\%(20\%) and 25\%(25\%) for the full, \tr and \tl production, respectively.

\section{Results}
\label{sec:results}
\begin{figure}[htbp!]
\centering  
\includegraphics[scale = 0.8]{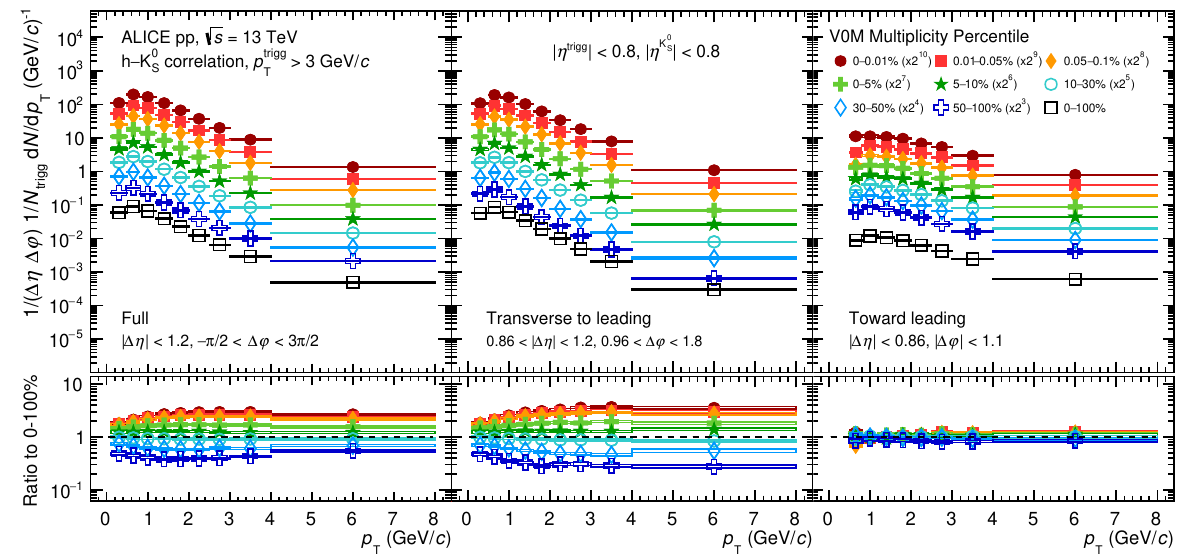}
\caption{Transverse momentum distributions of \kzero 
per unit \dEtadPhi area 
in \ppmb. The left, central and right panels refer to full, \tr and \tl production, respectively. Different colours refer to different multiplicity classes selected using the V0 detector, as indicated in the legend. 
The spectra are scaled by different factors to improve the visibility.
The bottom panels display the ratios to the spectra measured in the 0--100\% multiplicity class.
The statistical errors are represented by the error bars, the systematic uncertainties by the empty boxes.
Error bars are smaller than the marker size and are therefore not visible.}
\label{fig:PtSpectraK0s13TeV}   
\end{figure}

\begin{figure}[htbp!]
\centering  
\includegraphics[scale = 0.8]{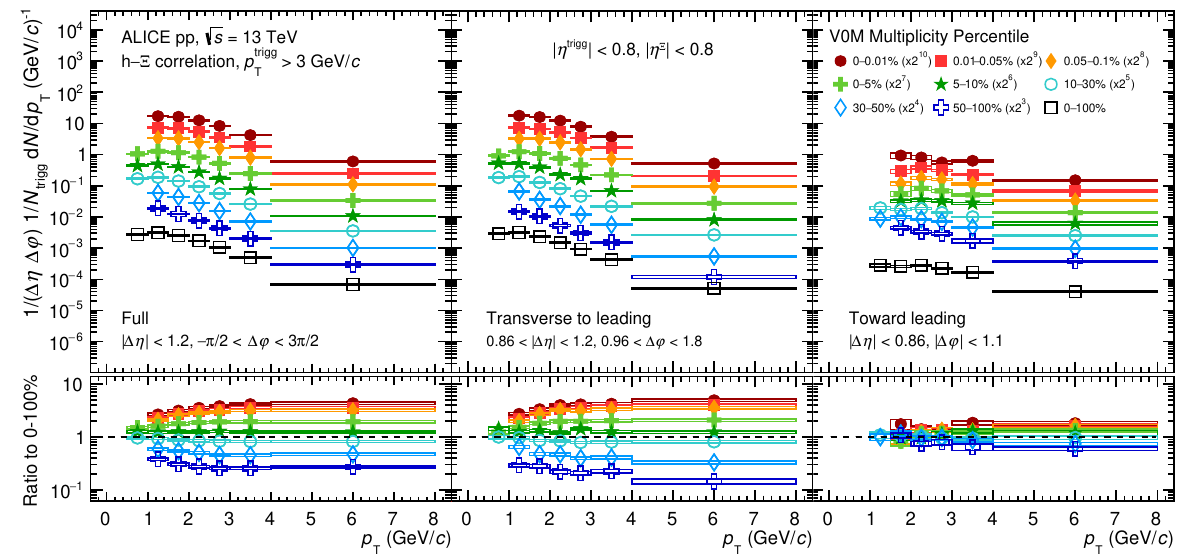}
\caption{Transverse momentum distributions of \Xis
per unit \dEtadPhi area 
in \ppmb. The left, central and right panels refer to full, transverse-to-leading and \tl production, respectively. Different colours refer to different multiplicity classes selected using the V0 detector, as indicated in the legend. 
The spectra are scaled by different factors to improve the visibility.
The bottom panels display the ratios to the spectra measured in the 0--100\% multiplicity class.
The statistical errors are represented by the error bars, the systematic uncertainties by the empty boxes.
Error bars are smaller than the marker size and are therefore not visible.
}
\label{fig:PtSpectraXi13TeV}   
\end{figure}

\begin{figure}[htbp!]
\centering  
\includegraphics[scale = 0.8]{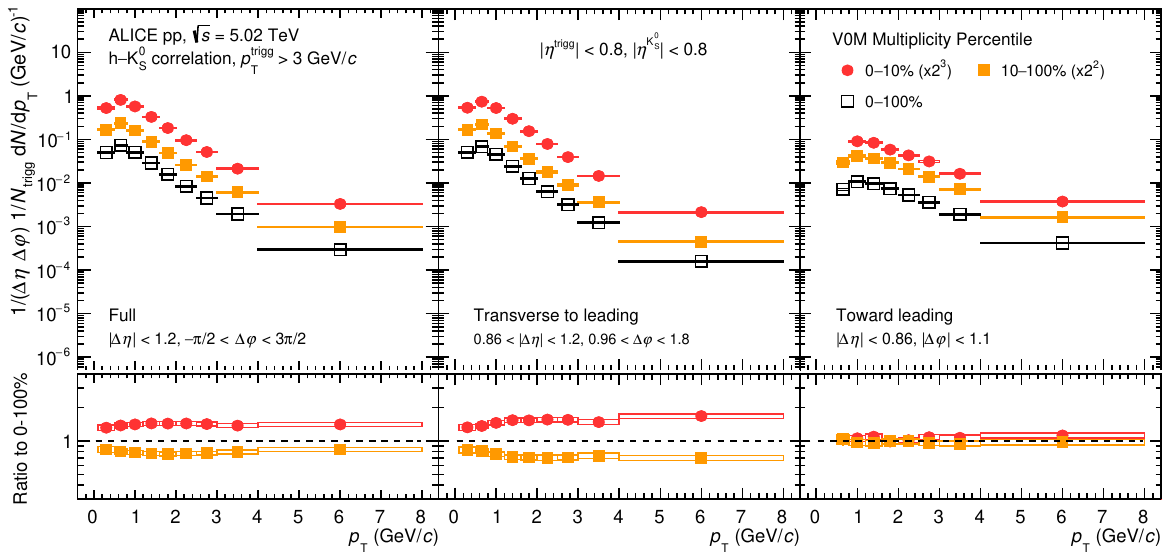}
\caption{Transverse momentum distributions of \kzero 
per unit \dEtadPhi area 
in \ppfi. The left, central and right panels refer to full, transverse-to-leading and \tl production, respectively. Different colours refer to different multiplicity classes selected using the V0 detector, as indicated in the legend. 
The spectra are scaled by different factors to improve the visibility.
The bottom panels display the ratios to the spectra measured in the 0--100\% multiplicity class.
The statistical errors are represented by the error bars, the systematic uncertainties by the empty boxes.
Error bars are smaller than the marker size and are therefore not visible.
}
\label{fig:PtSpectraK0s5TeV}   
\end{figure}

\begin{figure}[htbp!]
\centering  
\includegraphics[scale = 0.8]{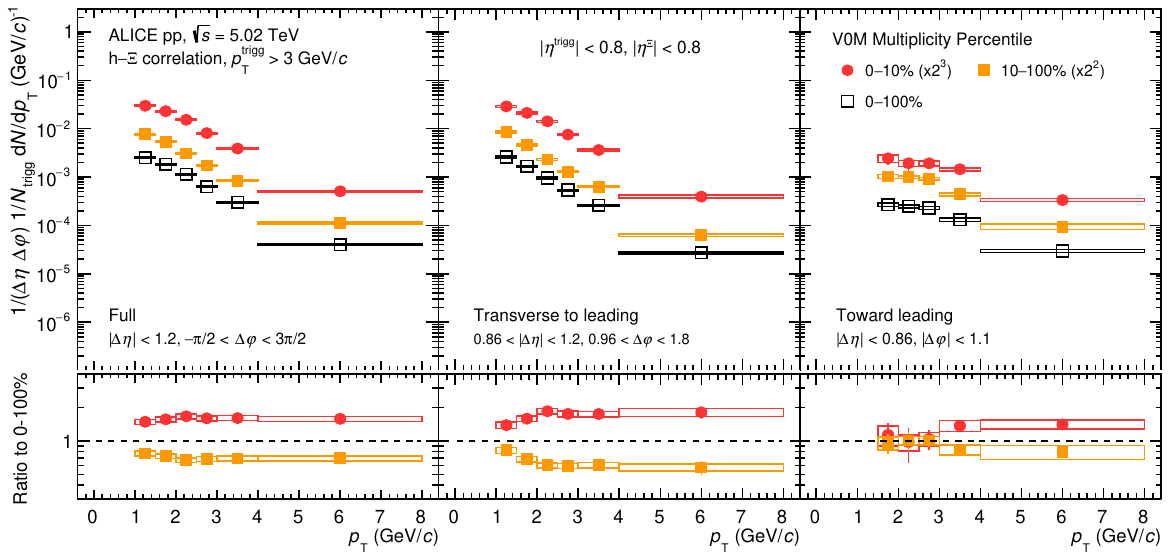}
\caption{Transverse momentum distributions of \Xis 
per unit \dEtadPhi area 
in \ppfi. The left, central and right panels refer to full, transverse-to-leading and \tl production, respectively. Different colours refer to different multiplicity classes selected using the V0 detector, as indicated in the legend. 
The spectra are scaled by different factors to improve the visibility.
The bottom panels display the ratios to the spectra measured in the 0--100\% multiplicity class.
The statistical errors are represented by the error bars, the systematic uncertainties by the empty boxes.
Error bars are smaller than the marker size and are therefore not visible.}
\label{fig:PtSpectraXi5TeV}   
\end{figure}

The full, \tl and \tr \pt distributions of \kzero and \Xis per unit \dEtadPhi area are shown for the different multiplicity classes in Figs.~\ref{fig:PtSpectraK0s13TeV} and~\ref{fig:PtSpectraXi13TeV} for pp collisions at \thTeV and Figs.~\ref{fig:PtSpectraK0s5TeV} and~\ref{fig:PtSpectraXi5TeV} for pp collisions at \fiTeV. 
The bottom panels show the ratios 
to the spectra measured in the 0--100\% multiplicity class.
In all multiplicity classes and at both centre-of-mass energies, the \tl spectra (right panels) are harder, i.e. have a larger average \pt, than the \tr (central panels) and full (left panels) spectra, as expected from the fact that the production in the direction of the trigger particle is associated with hard scattering processes.
As shown in the bottom panels of Figs.~\ref{fig:PtSpectraK0s13TeV}--\ref{fig:PtSpectraXi5TeV}, the \tr and full \pt spectra increase with multiplicity in all \pt intervals, becoming harder as the multiplicity increases. 
This behaviour was already reported for strange hadron spectra measured inclusively, i.e. in all events, in Pb--Pb~\cite{ALICE:2013xmt}, p--Pb~\cite{ALICE:2015mpp} and pp collisions~\cite{ALICE:2018pal, ALICE:2019avo}. 
In Pb--Pb collisions this behaviour is more pronounced than in small collision systems and is interpreted as an indication of the presence of radial flow.
In contrast to the full and \tr spectra, the \tl spectra show a much smaller dependence on the multiplicity.
 
\begin{figure}[b]
\centering
{\includegraphics[scale=0.8]{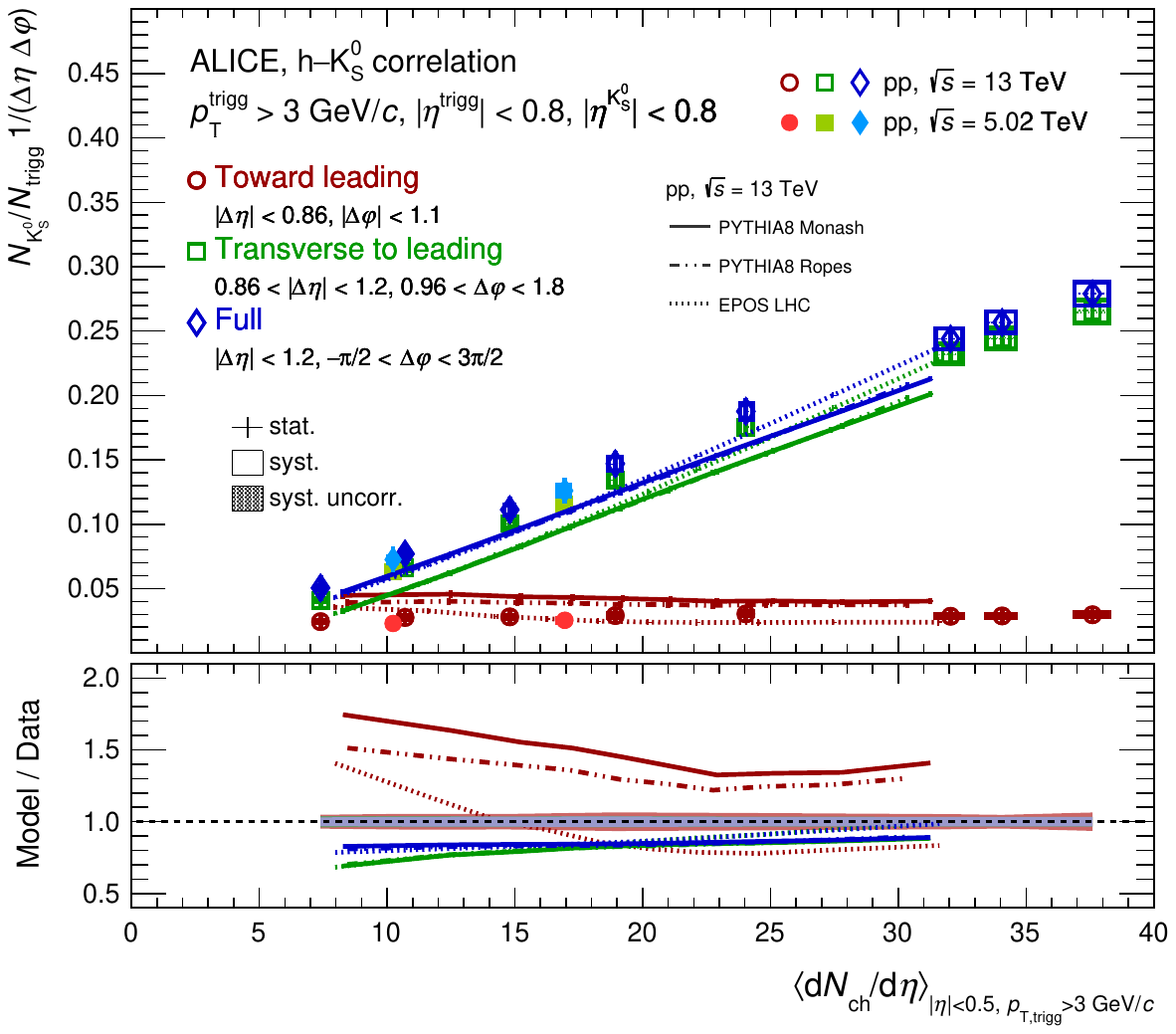}}
\caption{Upper panel: full (blue), \tr (green) and \tl (red) \kzero yields per unit \dEtadPhi area as a function of the charged-particle multiplicity measured in events with a trigger particle. The data points are drawn with markers, the model predictions with lines of different styles, as indicated in the legend. 
Statistical and systematic uncertainties of the data points are shown by error bars and empty boxes, respectively. Shadowed boxes 
 represent systematic uncertainties uncorrelated across multiplicity. The sum in quadrature of statistical and systematic uncertainties of the model predictions are shown by error bars, too small to be visible in the plot.
Bottom panel: ratio between the model predictions and the cubic spline fitted to the data points. The shaded band around one represents the sum in quadrature of the statistical and systematic uncertainties of the data points.}
\label{fig:YieldKs}   
\end{figure}

\begin{figure}[htbp!]
\centering  
{\includegraphics[scale=0.8]{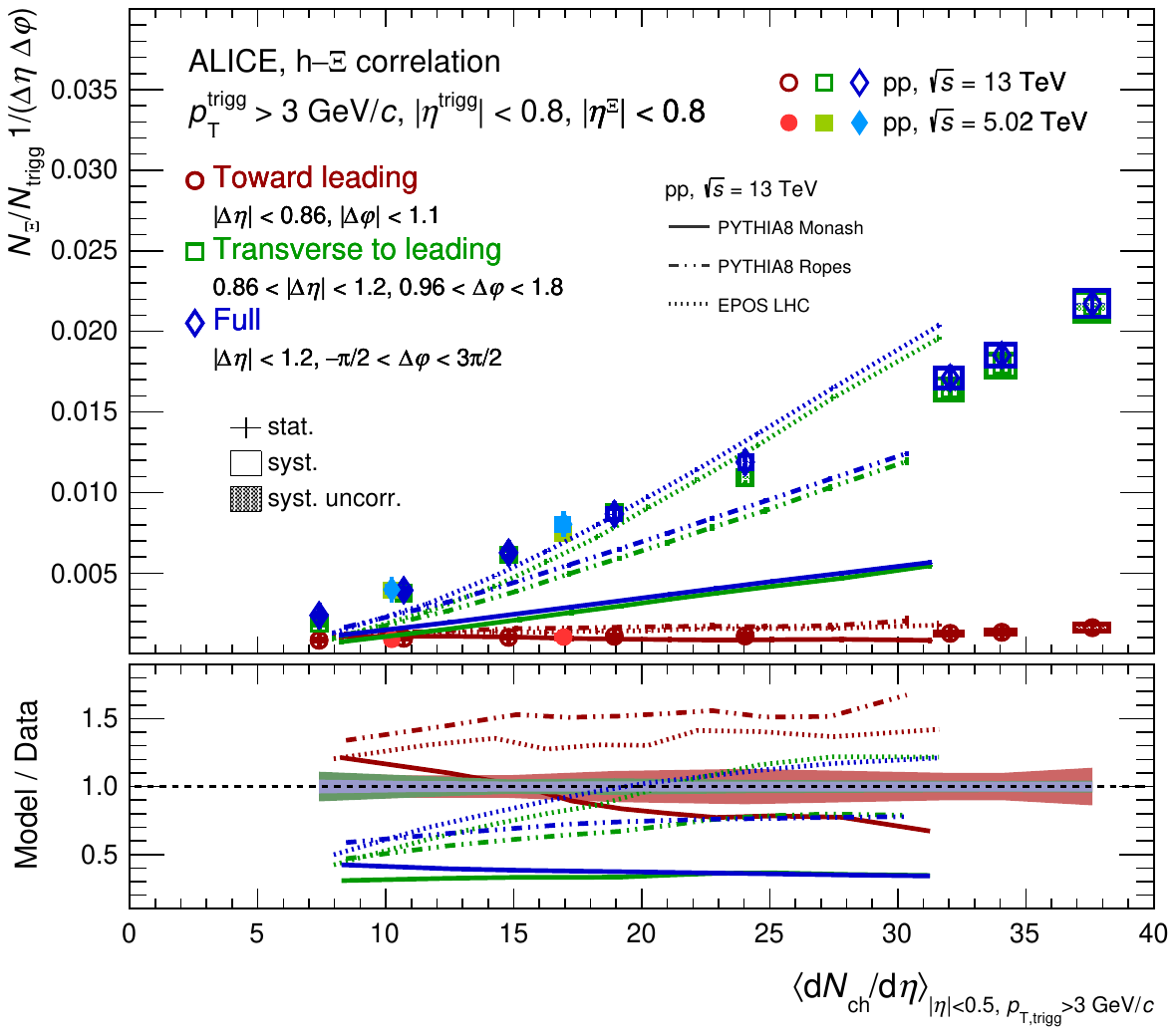}}
\caption{Upper panel: full (blue), \tr (green) and \tl (red) \Xis yields per unit \dEtadPhi area as a function of the charged-particle multiplicity measured in events with a trigger particle. The data points are drawn with markers, the model predictions with lines of different styles, as indicated in the legend. 
Statistical and systematic uncertainties of the data points are shown by error bars and empty boxes, respectively. Shadowed boxes represent systematic uncertainties uncorrelated across multiplicity. The sum in quadrature of statistical and systematic uncertainties of the model predictions are shown by error bars, too small to be visible in the plot.
Bottom panel: ratio between the model predictions and the cubic spline fitted to the data points. The shaded band around one represents the sum in quadrature of the statistical and systematic uncertainties of the data points.}
\label{fig:YieldXi}   
\end{figure}

\begin{figure}[b]
\centering
{\includegraphics[scale=0.8]{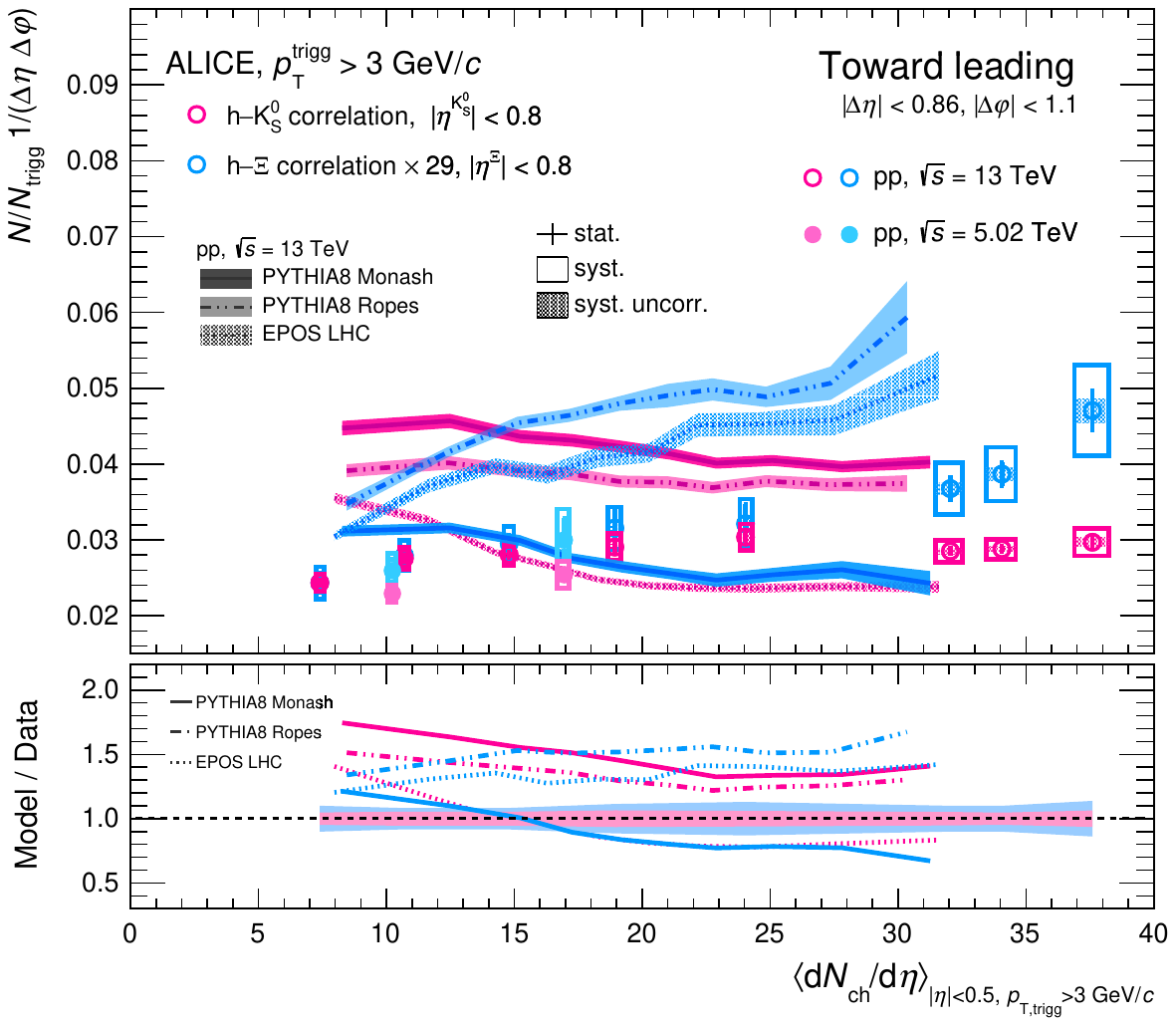}}
\caption{Upper panel: toward-leading \kzero (magenta) and \Xis (light blue) yields per unit \dEtadPhi area as a function of the charged-particle multiplicity measured in events with a trigger particle. The \Xis yields are scaled such that the lowest-multiplicity \Xis yield matches the \kzero one. The data points are drawn with markers, the model predictions with lines of different styles, as indicated in the legend. 
Statistical and systematic uncertainties of the data points are shown by error bars and empty boxes, respectively. Shadowed boxes 
 represent systematic uncertainties uncorrelated across multiplicity. 
The width of the bands represents the sum in quadrature of statistical and systematic uncertainties of the model predictions.
Bottom panel: ratio between the model predictions and the cubic spline fitted to the data points. The shaded band around one represents the sum in quadrature of the statistical and systematic uncertainties of the data points.}
\label{fig:YieldKsXiJet}   
\end{figure}

The full, \tr and \tl \pt-integrated yields of \kzero (\Xis) per unit \dEtadPhi area are shown in Fig.~\ref{fig:YieldKs} (\ref{fig:YieldXi}) as a function of the charged-particle multiplicity measured at midrapidity in events with a trigger particle \dNdetatrigg, in the following abbreviated with \dNdetaabbtrigg. 
The yields show no dependence on the centre-of-mass energy, as observed in previously published results~\cite{ALICE:2019avo}.
The full and \tr yields of both \kzero and \Xis increase with multiplicity faster than the \tl yields.
For better visibility, the \tl \pt-integrated yields of \kzero and \Xis per unit \dEtadPhi area are separately shown in Fig.~\ref{fig:YieldKsXiJet}, where the \Xis yields are scaled such that the lowest-multiplicity \Xis yield matches the \kzero one. 
Both the \kzero and \Xis yields are not compatible with a flat trend with multiplicity with a $5\sigma$ confidence level.
The relative increase of the \kzero yield from the lowest to the highest multiplicity is ($1.22\pm0.04$), where the uncertainty is given by the sum in quadrature of the statistical and the systematic uncertainty uncorrelated in multiplicity. The relative increase of the \Xis yield ($1.93\pm0.17$) is significantly larger than the \kzero one. 

The yields are compared with the predictions of three different phenomenological models, namely \textsc{Pythia}8.2 with the Monash 2013 tune~\cite{Sjostrand:2014zea}, \textsc{Pythia}8.2 with ropes~\cite{Bierlich:2014xba} and EPOS LHC~\cite{Pierog:2013ria}. 
\textsc{Pythia} is based on the Lund string hadronisation model~\cite{Andersson:1997xwk}.
As shown in Ref.~\cite{ALICE:2020nkc}, \textsc{Pythia}8.2 with the Monash 2013 tune cannot describe the strangeness enhancement with multiplicity in $\mathrm{INEL>0}$  pp collisions: it underestimates the ratios between strange hadron and pion yields and does not reproduce their increase with multiplicity.
The description is improved if overlapping strings are allowed to interact with each other, forming the so-called ``colour ropes"~\cite{Bierlich:2014xba}. 
Indeed, \textsc{Pythia}8 with colour ropes can qualitatively describe the strangeness enhancement with multiplicity in pp collisions, as shown in Ref.~\cite{ALICE:2020nkc}.
Finally, the EPOS LHC~\cite{Pierog:2013ria} event generator implements the core-corona model~\cite{PhysRevLett.98.152301}, according to which strings in a low-density area form the corona and hadronise normally, while strings in a high-density area form the core and undergo collective hadronisation. As shown in Ref.~\cite{ALICE:2016fzo}, EPOS LHC can reasonably well describe the $\mathrm{K^0_S}/\pi$ ratio measured in $\mathrm{INEL>0}$ pp collisions, while it overestimates the strangeness enhancement with multiplicity for the $\Lambda$, $\Xi^\pm$ and $\Omega^\pm$ baryons.

The bottom panels of Figs.~\ref{fig:YieldKs}-\ref{fig:YieldXi}-\ref{fig:YieldKsXiJet} display the ratios between the model predictions and the cubic splines fitted to the data points. 
Three sources of systematic uncertainty affecting the model predictions were considered: the choice of $\Delta\eta$ and $\Delta\varphi$ regions, which is evaluated as described in Section~\ref{sec:sys} for the data and is found not to be significant according to the Barlow criterion for both the \tl and the \tr production, and the extrapolation of the yields in the unmeasured \pt regions: $p_\mathrm{T} < 0.5(1.0)$~GeV/\textit{c} for \kzero (\Xis) \tl yield and $p_\mathrm{T} < 0.5$~GeV/\textit{c} for \Xis full and \tr yields. 
All the models underestimate the full and the \tr \kzero yields (Fig.~\ref{fig:YieldKs}). 
The underestimation is more significant at low multiplicity ($\dNdetaabbtrigg \sim10$), where all models underestimate the yields by about 30\%. At high multiplicity ($\dNdetaabbtrigg \sim30$), both \textsc{Pythia}8 implementations underestimate the yields by about 15\%, while EPOS LHC predicts values compatible with the measured ones. 
The increase with multiplicity of the \tl yield of \kzero (Fig.~\ref{fig:YieldKsXiJet}) is not reproduced by any of the three models: both \textsc{Pythia}8 implementations overestimate the yields and show a hint of decrease with multiplicity, whereas EPOS LHC predicts a decrease of the \tl yield with multiplicity.
The deviation of the models from the full and \tr \Xis yields (Fig.~\ref{fig:YieldXi}) is larger than the deviation from those of the \kzero (Fig.~\ref{fig:YieldKs}).
Both \textsc{Pythia}8 implementations underestimate the yields: \textsc{Pythia}8 Monash underestimates them by approximately 70\% over the whole multiplicity interval, whereas \textsc{Pythia}8 with ropes underestimates them by about 50\% at low multiplicity and 20\% at high multiplicity.
EPOS LHC underestimates the yield at low multiplicity by about 50\% and overestimates it by about 20\% at high multiplicity, predicting an increase of the \tr and full yields with multiplicity larger than the one observed in the data.
The increase with multiplicity of the \Xis \tl yield (Fig.~\ref{fig:YieldKsXiJet}) is not described by \textsc{Pythia}8 Monash, which predicts a nearly flat trend with multiplicity.  On the contrary, \textsc{Pythia}8 with ropes and EPOS LHC can qualitatively reproduce the increasing trend. These models, however, overestimate the \tl yields over the whole multiplicity interval. 

\begin{figure}[htbp!]
\centering  
{\includegraphics[scale=0.7]{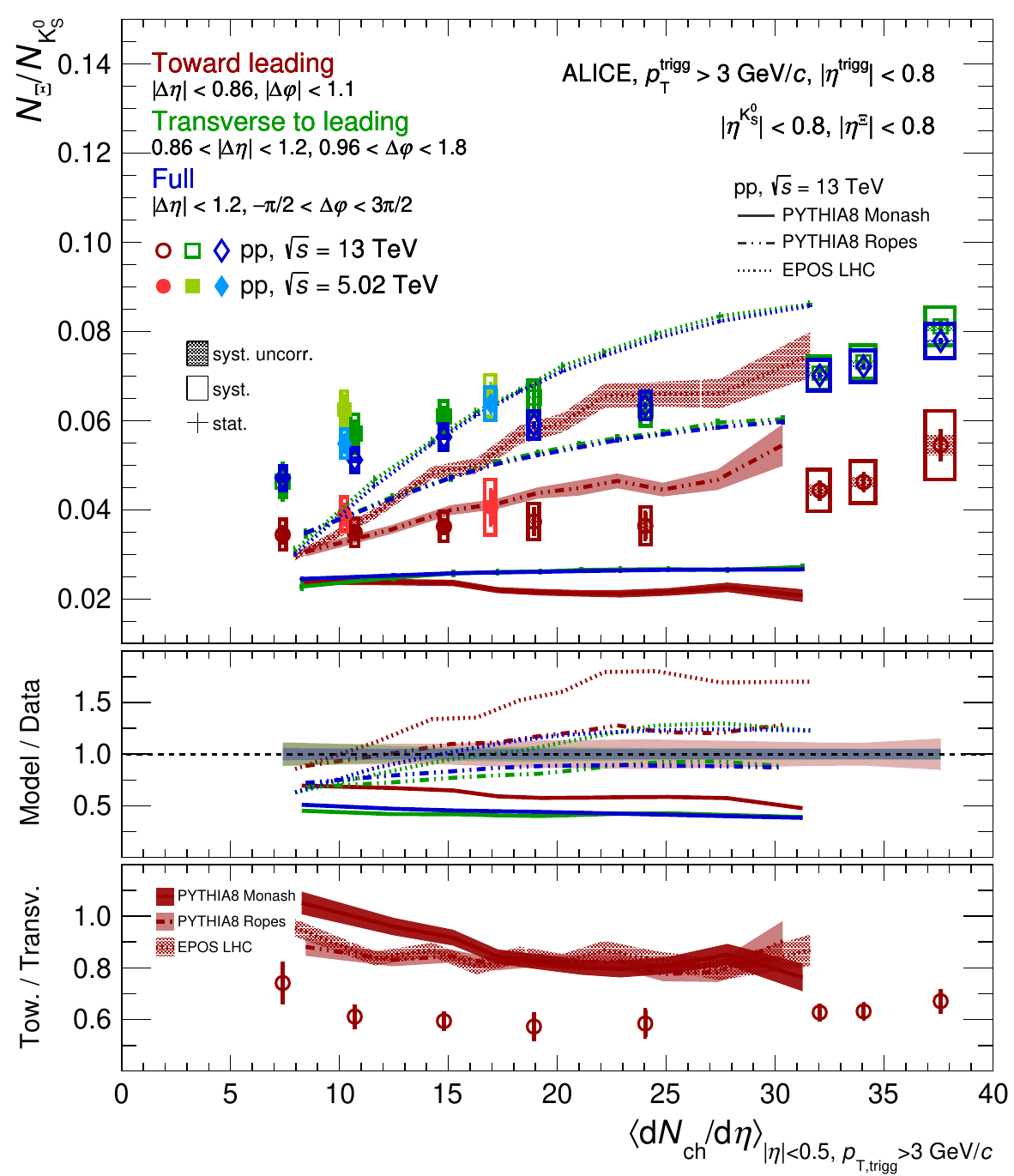}}
\caption{
Top panel: full (blue), \tr (green) and \tl (red) \Xis/\kzero yield ratios as a function of the charged-particle multiplicity measured at midrapidity in events with a trigger particle. 
The data points are drawn with markers,
and their statistical and systematic uncertainties are shown by error bars and empty boxes, respectively. Shadowed boxes represent systematic uncertainties uncorrelated across multiplicity.
The model predictions are drawn with lines of different styles. The width of the bands represents the sum in quadrature of statistical and systematic uncertainties of the model predictions, and is visible only for \tl production.
Central panel: ratio between the model predictions and the cubic spline fitted to the data points. The shaded band around unity represents the sum in quadrature of the statistical and systematic uncertainties of the data points.
Bottom panel: double ratio between the \tl and the \tr \Xis/\kzero ratios. 
}
\label{fig:RatioXiK0s}   
\end{figure}

The ratios between \Xis and \kzero yields as a function of \dNdetaabbtrigg are shown in the top panel of Fig.~\ref{fig:RatioXiK0s}, together with the model predictions. 
In the data, the ratio of full yields increases with multiplicity: this could be related to the larger strangeness content of the \Xis with respect to the \kzero. Indeed, the enhanced production of strange hadrons with increasing multiplicity was observed to be higher for particles with larger strangeness content~\cite{ALICE:2016fzo}.
The ratio of \tr yields increases with the multiplicity by a factor $(1.75\pm0.16)$, with the error given by the sum in quadrature of the statistical and systematic uncertainty uncorrelated across multiplicity. It is compatible with the ratio of full yields, because the full yield is dominated by \tr production, as shown in Figs.~\ref{fig:YieldKs} and~\ref{fig:YieldXi}. 
Also the \tl ratio increases with multiplicity: a flat behaviour with multiplicity is excluded since a zero-degree polynomial is not able to describe the ratio within the uncertainties uncorrelated across multiplicity. 
The increase of the \tl ratio from the lowest to the highest multiplicity interval equals a factor ($1.58\pm0.15$).
As shown by the double ratio between the \tl and the \tr \Xis/\kzero ratios displayed in the bottom panel of Fig.~\ref{fig:RatioXiK0s}, the \tl ratio is approximately 40\% smaller than the \tr ratio, suggesting that the production of \Xis with respect to \kzero is favoured in \tr processes over the whole multiplicity interval where the measurement was performed. 
The double ratio is well described by a zero-degree polynomial with a $\chi^2/\mathrm{ndf} = 6.6/7$, indicating that the \tr and \tl yield ratios increase with multiplicity in a similar way.

The central panel of Fig.~\ref{fig:RatioXiK0s} displays the ratio between the model predictions and the data points.
\textsc{Pythia}8 Monash underestimates the ratios in the whole multiplicity interval, due to the large underestimation of the full and \tr \Xis yields and of the overestimation of the \kzero \tl yield. 
The \tl ratio does not describe the increase observed in the data. 
The full and \tr ratios show instead an increase with multiplicity, which is smaller than the one observed in the data, as suggested by the decrease of the model over data ratio from about 0.5 at low multiplicity to about 0.4 at high multiplicity.
\textsc{Pythia}8 with ropes can qualitatively describe the increase of the ratios with multiplicity observed in the data. However, the full and \tr ratios are underestimated in the whole multiplicity interval, and particularly at low multiplicity, where the underestimation of the \Xis yields is larger. The \tl ratio is in qualitative agreement with the data, but its increase with multiplicity is slightly overestimated: this fair agreement is resulting from the overestimation of both the \kzero and the \Xis \tl yields.
Finally, EPOS LHC overestimates the increase with multiplicity of the full and \tr ratios, as a consequence of the overestimation of the increase with multiplicity of \Xis yields. In particular, the full and \tr ratios are underestimated by about 30\% at low multiplicity and overestimated by about 20\% at high multiplicity.
The \tl ratio is instead overestimated in the whole multiplicity interval, mainly as a consequence of the overestimation of the \Xis \tl yield. Moreover, its increase with multiplicity is larger than the one observed in the data because the \kzero \tl yields predicted by EPOS LHC decrease with multiplicity.
As shown in the bottom panel of Fig.~\ref{fig:RatioXiK0s}, the three models predict a larger double ratio than the one measured in the data, i.e. they overestimate the \tl \Xis/\kzero production with respect to the \tr one. 
The double ratios predicted by \textsc{Pythia}8 with ropes and EPOS LHC are smaller than unity and can be described with a zero-degree polynomial. On the contrary, the double ratio predicted by \textsc{Pythia}8 Monash is compatible with one in the lowest multiplicity class and decreases to about 0.8 in the highest multiplicity class.

The comparison of the Monte Carlo model predictions with the data suggests that none of the considered models describes strange hadron production in hard scattering processes or in the underlying event.

\section{Summary and outlook}
\label{sec:outlook}
The production of \kzero and \Xis in \ppfi and at \thTeV was measured in the direction of the highest-\pt charged particle (trigger particle) and in the direction transverse to it. 
The \tl \pt spectra are harder than the \tr ones, as expected from the fact that the production in the direction of the trigger particle is associated with hard scattering processes, whereas the production in the transverse-to-leading direction is related to the underlying event. 

The full \pt-integrated yields per unit \dEtadPhi of \kzero and \Xis are dominated by \tr production and increase with the multiplicity of charged particles produced at midrapidity. The \tl yields show instead a milder dependence on the multiplicity, indicating that the contribution of \tr processes relative to \tl ones increases with the multiplicity. The \kzero and \Xis yields do not show any significant centre-of-mass energy dependence.

The ratio between the \Xis and the \kzero yields provides insight into the strangeness enhancement effect, since the strangeness content of the \Xis ($|S|$=2) is larger than the \kzero one ($|S|$=1). 
Both the \tr and the \tl \Xis/\kzero yield ratios increase with the multiplicity of charged particles. 
The \tr ratio is larger than the \tl one, suggesting that the relative production of \Xis with respect to \kzero is favoured in underlying event processes. 

None of the considered models, namely \textsc{Pythia}8 Monash tune, \textsc{Pythia}8 with ropes and EPOS LHC, can quantitatively describe the \tr and \tl yields of \kzero and \Xis. 
Both \textsc{Pythia}8 implementations underestimate the full and the \tr \kzero and \Xis yields, with the largest underestimation observed for the \Xis yields.  
The increase of the full and \tr \Xis yields with multiplicity is overestimated by both \textsc{Pythia}8 with ropes and EPOS LHC, leading to an  overestimation of the increase of the full and \tr \Xis/\kzero yield ratios with multiplicity. 
The increase of the \tl yield of \kzero with multiplicity is not reproduced by any of the three models.  
On the contrary, the increase of the \Xis \tl yield with multiplicity is qualitatively reproduced by \textsc{Pythia}8 with ropes and EPOS LHC, while \textsc{Pythia}8 Monash predicts a flat trend with multiplicity. 
Overall, the comparison with the data indicates that the strange hadron production associated with both hard scattering processes and the underlying event is not properly described by any of the considered models.
Additionally, other models such as the most recent implementation of the core-corona approach EPOS4~\cite{Werner:2023jps} could be tested.

Further investigation of the origin of the enhanced production of strange hadrons in high-multiplicity pp collisions with respect to low-multiplicity ones will be possible thanks to the huge sample of pp collisions that is being collected during the ongoing Run~3, which is expected to be three orders of magnitude larger than the Run~2 one. 
With Run~3 data, measuring the \tl and \tr yields of the triple-strange baryon \Oms as a function of the charged-particle multiplicity will become feasible. Additionally, it will be possible to study the dependence of the \tl and \tr \Xis yields on the minimum \pt of the trigger particle, with higher \pt thresholds reducing the contamination from particles not originating from hard-scattering events.
These studies will help improve the current understanding of strange hadron production mechanisms.


\newenvironment{acknowledgement}{\relax}{\relax}
\begin{acknowledgement}
\section*{Acknowledgements}
\input{fa_2024-04-23_Opt_C.tex}
\end{acknowledgement}

\bibliographystyle{utphys}   
\bibliography{bibliography}

\newpage
\appendix

%
%

\section{The ALICE Collaboration}
\label{app:collab}
\input{Alice_Authorlist_2024-04-23_Opt_C.tex}
\end{document}

%% file: commands.tex
%

\newcommand{\pp}           {pp\xspace}
\newcommand{\ppbar}        {\mbox{$\mathrm {p\overline{p}}$}\xspace}
\newcommand{\XeXe}         {\mbox{Xe--Xe}\xspace}
\newcommand{\PbPb}         {\mbox{Pb--Pb}\xspace}
\newcommand{\pA}           {\mbox{pA}\xspace}
\newcommand{\pPb}          {\mbox{p--Pb}\xspace}
\newcommand{\AuAu}         {\mbox{Au--Au}\xspace}
\newcommand{\dAu}          {\mbox{d--Au}\xspace}

\newcommand{\s}            {\ensuremath{\sqrt{s}}\xspace}
\newcommand{\snn}          {\ensuremath{\sqrt{s_{\mathrm{NN}}}}\xspace}
\newcommand{\pt}           {\ensuremath{p_{\rm T}}\xspace}
\newcommand{\meanpt}       {$\langle p_{\mathrm{T}}\rangle$\xspace}
\newcommand{\ycms}         {\ensuremath{y_{\rm CMS}}\xspace}
\newcommand{\ylab}         {\ensuremath{y_{\rm lab}}\xspace}
\newcommand{\etarange}[1]  {\mbox{$\left | \eta \right |~<~#1$}}
\newcommand{\yrange}[1]    {\mbox{$\left | y \right |~<~#1$}}
\newcommand{\dndy}         {\ensuremath{\mathrm{d}N_\mathrm{ch}/\mathrm{d}y}\xspace}
\newcommand{\dndeta}       {\ensuremath{\mathrm{d}N_\mathrm{ch}/\mathrm{d}\eta}\xspace}
\newcommand{\avdndeta}     {\ensuremath{\langle\dndeta\rangle}\xspace}
\newcommand{\dNdy}         {\ensuremath{\mathrm{d}N_\mathrm{ch}/\mathrm{d}y}\xspace}
\newcommand{\Npart}        {\ensuremath{N_\mathrm{part}}\xspace}
\newcommand{\Ncoll}        {\ensuremath{N_\mathrm{coll}}\xspace}
\newcommand{\dEdx}         {\ensuremath{\textrm{d}E/\textrm{d}x}\xspace}
\newcommand{\RpPb}         {\ensuremath{R_{\rm pPb}}\xspace}

\newcommand{\nineH}        {$\sqrt{s}~=~0.9$~Te\kern-.1emV\xspace}
\newcommand{\seven}        {$\sqrt{s}~=~7$~Te\kern-.1emV\xspace}
\newcommand{\twoH}         {$\sqrt{s}~=~0.2$~Te\kern-.1emV\xspace}
\newcommand{\twosevensix}  {$\sqrt{s}~=~2.76$~Te\kern-.1emV\xspace}
\newcommand{\five}         {$\sqrt{s}~=~5.02$~Te\kern-.1emV\xspace}
\newcommand{\twosevensixnn}{$\sqrt{s_{\mathrm{NN}}}~=~2.76$~Te\kern-.1emV\xspace}
\newcommand{\fivenn}       {$\sqrt{s_{\mathrm{NN}}}~=~5.02$~Te\kern-.1emV\xspace}
\newcommand{\LT}           {L{\'e}vy-Tsallis\xspace}
\newcommand{\GeVc}         {Ge\kern-.1emV/$c$\xspace}
\newcommand{\MeVc}         {Me\kern-.1emV/$c$\xspace}
\newcommand{\TeV}          {Te\kern-.1emV\xspace}
\newcommand{\GeV}          {Ge\kern-.1emV\xspace}
\newcommand{\MeV}          {Me\kern-.1emV\xspace}
\newcommand{\GeVmass}      {Ge\kern-.2emV/$c^2$\xspace}
\newcommand{\MeVmass}      {Me\kern-.2emV/$c^2$\xspace}
\newcommand{\lumi}         {\ensuremath{\mathcal{L}}\xspace}

\newcommand{\ITS}          {\rm{ITS}\xspace}
\newcommand{\TOF}          {\rm{TOF}\xspace}
\newcommand{\ZDC}          {\rm{ZDC}\xspace}
\newcommand{\ZDCs}         {\rm{ZDCs}\xspace}
\newcommand{\ZNA}          {\rm{ZNA}\xspace}
\newcommand{\ZNC}          {\rm{ZNC}\xspace}
\newcommand{\SPD}          {\rm{SPD}\xspace}
\newcommand{\SDD}          {\rm{SDD}\xspace}
\newcommand{\SSD}          {\rm{SSD}\xspace}
\newcommand{\TPC}          {\rm{TPC}\xspace}
\newcommand{\TRD}          {\rm{TRD}\xspace}
\newcommand{\VZERO}        {\rm{V0}\xspace}
\newcommand{\VZEROA}       {\rm{V0A}\xspace}
\newcommand{\VZEROC}       {\rm{V0C}\xspace}
\newcommand{\Vdecay} 	   {\ensuremath{V^{0}}\xspace}

\newcommand{\ee}           {\ensuremath{e^{+}e^{-}}} 
\newcommand{\pip}          {\ensuremath{\pi^{+}}\xspace}
\newcommand{\pim}          {\ensuremath{\pi^{-}}\xspace}
\newcommand{\kap}          {\ensuremath{\rm{K}^{+}}\xspace}
\newcommand{\kam}          {\ensuremath{\rm{K}^{-}}\xspace}
\newcommand{\pbar}         {\ensuremath{\rm\overline{p}}\xspace}
\newcommand{\kzero}        {\ensuremath{{\rm K}^{0}_{\rm{S}}}\xspace}
\newcommand{\lmb}          {\ensuremath{\Lambda}\xspace}
\newcommand{\almb}         {\ensuremath{\overline{\Lambda}}\xspace}
\newcommand{\Om}           {\ensuremath{\Omega^-}\xspace}
\newcommand{\Mo}           {\ensuremath{\overline{\Omega}^+}\xspace}
\newcommand{\X}            {\ensuremath{\Xi^-}\xspace}
\newcommand{\Ix}           {\ensuremath{\overline{\Xi}^+}\xspace}
\newcommand{\Xis}          {\ensuremath{\Xi^{\pm}}\xspace}
\newcommand{\Oms}          {\ensuremath{\Omega^{\pm}}\xspace}
\newcommand{\degree}       {\ensuremath{^{\rm o}}\xspace}

\newcommand{\Xiplus}    {\ensuremath{\Xi^{+}}\xspace}
\newcommand{\Ximinus}    {\ensuremath{\Xi^{-}}\xspace}
\newcommand{\tl}{\mbox{toward-leading}\xspace}

\newcommand{\tr}{\mbox{transverse-to-leading}\xspace}

\newcommand{\hKs}{h-\ensuremath{\rm{K}^0_{\rm{S}}}\xspace}

\newcommand{\hXi}{h-\ensuremath{\Xi^{\pm}\xspace}}

\newcommand{\dNdetatrigg}     {\ensuremath{\langle\dndeta\rangle_{|\eta|< 0.5,\ p_{\rm{T, trigg}}>3\ \mathrm{GeV}/c}}\xspace}

\newcommand{\dNdetaabb}
{\ensuremath{\langle \mathrm{d}N/\mathrm{d}\eta \rangle}\xspace}

\newcommand{\dNdetaabbtrigg}
{\ensuremath{\langle \mathrm{d}N/\mathrm{d}\eta \rangle _\mathrm {trigg}}\xspace}

\newcommand{\ppmb}{pp collisions at \mbox{\ensuremath{\sqrt{s}=13}~TeV}\xspace}
\newcommand{\thTeV}{\mbox{\ensuremath{\sqrt{s}=13}~TeV}\xspace}

\newcommand{\ppfi}{pp collisions at \mbox{\ensuremath{\sqrt{s}=5.02}~TeV}\xspace}
\newcommand{\fiTeV}{\mbox{\ensuremath{\sqrt{s}=5.02}~TeV}\xspace}

\newcommand{\dPhi}{\ensuremath{\Delta\varphi}\xspace}

\newcommand{\dEta}{\ensuremath{\Delta\eta}\xspace}

\newcommand{\dEtadPhi}{\ensuremath{\Delta\eta\Delta\varphi}\xspace}

%% file: fa_2024-04-23_Opt_C.tex

The ALICE Collaboration would like to thank all its engineers and technicians for their invaluable contributions to the construction of the experiment and the CERN accelerator teams for the outstanding performance of the LHC complex.
The ALICE Collaboration gratefully acknowledges the resources and support provided by all Grid centres and the Worldwide LHC Computing Grid (WLCG) collaboration.
The ALICE Collaboration acknowledges the following funding agencies for their support in building and running the ALICE detector:
A. I. Alikhanyan National Science Laboratory (Yerevan Physics Institute) Foundation (ANSL), State Committee of Science and World Federation of Scientists (WFS), Armenia;
Austrian Academy of Sciences, Austrian Science Fund (FWF): [M 2467-N36] and Nationalstiftung f\"{u}r Forschung, Technologie und Entwicklung, Austria;
Ministry of Communications and High Technologies, National Nuclear Research Center, Azerbaijan;
Conselho Nacional de Desenvolvimento Cient\'{\i}fico e Tecnol\'{o}gico (CNPq), Financiadora de Estudos e Projetos (Finep), Funda\c{c}\~{a}o de Amparo \`{a} Pesquisa do Estado de S\~{a}o Paulo (FAPESP) and Universidade Federal do Rio Grande do Sul (UFRGS), Brazil;
Bulgarian Ministry of Education and Science, within the National Roadmap for Research Infrastructures 2020-2027 (object CERN), Bulgaria;
Ministry of Education of China (MOEC) , Ministry of Science \& Technology of China (MSTC) and National Natural Science Foundation of China (NSFC), China;
Ministry of Science and Education and Croatian Science Foundation, Croatia;
Centro de Aplicaciones Tecnol\'{o}gicas y Desarrollo Nuclear (CEADEN), Cubaenerg\'{\i}a, Cuba;
Ministry of Education, Youth and Sports of the Czech Republic, Czech Republic;
The Danish Council for Independent Research | Natural Sciences, the VILLUM FONDEN and Danish National Research Foundation (DNRF), Denmark;
Helsinki Institute of Physics (HIP), Finland;
Commissariat \`{a} l'Energie Atomique (CEA) and Institut National de Physique Nucl\'{e}aire et de Physique des Particules (IN2P3) and Centre National de la Recherche Scientifique (CNRS), France;
Bundesministerium f\"{u}r Bildung und Forschung (BMBF) and GSI Helmholtzzentrum f\"{u}r Schwerionenforschung GmbH, Germany;
General Secretariat for Research and Technology, Ministry of Education, Research and Religions, Greece;
National Research, Development and Innovation Office, Hungary;
Department of Atomic Energy Government of India (DAE), Department of Science and Technology, Government of India (DST), University Grants Commission, Government of India (UGC) and Council of Scientific and Industrial Research (CSIR), India;
National Research and Innovation Agency - BRIN, Indonesia;
Istituto Nazionale di Fisica Nucleare (INFN), Italy;
Japanese Ministry of Education, Culture, Sports, Science and Technology (MEXT) and Japan Society for the Promotion of Science (JSPS) KAKENHI, Japan;
Consejo Nacional de Ciencia (CONACYT) y Tecnolog\'{i}a, through Fondo de Cooperaci\'{o}n Internacional en Ciencia y Tecnolog\'{i}a (FONCICYT) and Direcci\'{o}n General de Asuntos del Personal Academico (DGAPA), Mexico;
Nederlandse Organisatie voor Wetenschappelijk Onderzoek (NWO), Netherlands;
The Research Council of Norway, Norway;
Pontificia Universidad Cat\'{o}lica del Per\'{u}, Peru;
Ministry of Science and Higher Education, National Science Centre and WUT ID-UB, Poland;
Korea Institute of Science and Technology Information and National Research Foundation of Korea (NRF), Republic of Korea;
Ministry of Education and Scientific Research, Institute of Atomic Physics, Ministry of Research and Innovation and Institute of Atomic Physics and Universitatea Nationala de Stiinta si Tehnologie Politehnica Bucuresti, Romania;
Ministry of Education, Science, Research and Sport of the Slovak Republic, Slovakia;
National Research Foundation of South Africa, South Africa;
Swedish Research Council (VR) and Knut \& Alice Wallenberg Foundation (KAW), Sweden;
European Organization for Nuclear Research, Switzerland;
Suranaree University of Technology (SUT), National Science and Technology Development Agency (NSTDA) and National Science, Research and Innovation Fund (NSRF via PMU-B B05F650021), Thailand;
Turkish Energy, Nuclear and Mineral Research Agency (TENMAK), Turkey;
National Academy of  Sciences of Ukraine, Ukraine;
Science and Technology Facilities Council (STFC), United Kingdom;
National Science Foundation of the United States of America (NSF) and United States Department of Energy, Office of Nuclear Physics (DOE NP), United States of America.
In addition, individual groups or members have received support from:
Czech Science Foundation (grant no. 23-07499S), Czech Republic;
European Research Council (grant no. 950692), European Union;
ICSC - Centro Nazionale di Ricerca in High Performance Computing, Big Data and Quantum Computing, European Union - NextGenerationEU;
Academy of Finland (Center of Excellence in Quark Matter) (grant nos. 346327, 346328), Finland.

%% file: Alice_Authorlist_2024-04-23_Opt_C.tex
\begin{flushleft} 
\small

S.~Acharya\,\orcidlink{0000-0002-9213-5329}\,$^{\rm 127}$, 
D.~Adamov\'{a}\,\orcidlink{0000-0002-0504-7428}\,$^{\rm 86}$, 
A.~Agarwal$^{\rm 135}$, 
G.~Aglieri Rinella\,\orcidlink{0000-0002-9611-3696}\,$^{\rm 32}$, 
L.~Aglietta\,\orcidlink{0009-0003-0763-6802}\,$^{\rm 24}$, 
M.~Agnello\,\orcidlink{0000-0002-0760-5075}\,$^{\rm 29}$, 
N.~Agrawal\,\orcidlink{0000-0003-0348-9836}\,$^{\rm 25}$, 
Z.~Ahammed\,\orcidlink{0000-0001-5241-7412}\,$^{\rm 135}$, 
S.~Ahmad\,\orcidlink{0000-0003-0497-5705}\,$^{\rm 15}$, 
S.U.~Ahn\,\orcidlink{0000-0001-8847-489X}\,$^{\rm 71}$
% ...

\section*{Affiliation Notes}

$^{\rm I}$ Deceased\\
$^{\rm II}$ Also at: Max-Planck-Institut fur Physik, Munich, Germany\\
$^{\rm III}$ Also at: Italian National Agency for New Technologies, Energy and Sustainable Economic Development (ENEA), Bologna, Italy\\
$^{\rm IV}$ Also at: Dipartimento DET del Politecnico di Torino, Turin, Italy\\
$^{\rm V}$ Also at: Yildiz Technical University, Istanbul, T\"{u}rkiye\\
$^{\rm VI}$ Also at: Department of Applied Physics, Aligarh Muslim University, Aligarh, India\\
$^{\rm VII}$ Also at: Institute of Theoretical Physics, University of Wroclaw, Poland\\
$^{\rm VIII}$ Also at: An institution covered by a cooperation agreement with CERN\\

\section*{Collaboration Institutes}

$^{1}$ A.I. Alikhanyan National Science Laboratory (Yerevan Physics Institute) Foundation, Yerevan, Armenia\\
$^{2}$ AGH University of Krakow, Cracow, Poland\\
$^{3}$ Bogolyubov Institute for Theoretical Physics, National Academy of Sciences of Ukraine, Kiev, Ukraine\\
$^{4}$ Bose Institute, Department of Physics  and Centre for Astroparticle Physics and Space Science (CAPSS), Kolkata, India\\
$^{5}$ California Polytechnic State University, San Luis Obispo, California, United States\\
$^{6}$ Central China Normal University, Wuhan, China\\
$^{7}$ Centro de Aplicaciones Tecnol\'{o}gicas y Desarrollo Nuclear (CEADEN), Havana, Cuba\\
$^{8}$ Centro de Investigaci\'{o}n y de Estudios Avanzados (CINVESTAV), Mexico City and M\'{e}rida, Mexico\\
$^{9}$ Chicago State University, Chicago, Illinois, United States\\
$^{10}$ China Institute of Atomic Energy, Beijing, China\\
$^{11}$ China University of Geosciences, Wuhan, China\\
$^{12}$ Chungbuk National University, Cheongju, Republic of Korea\\
$^{13}$ Comenius University Bratislava, Faculty of Mathematics, Physics and Informatics, Bratislava, Slovak Republic\\
$^{14}$ Creighton University, Omaha, Nebraska, United States\\
$^{15}$ Department of Physics, Aligarh Muslim University, Aligarh, India\\
$^{16}$ Department of Physics, Pusan National University, Pusan, Republic of Korea\\
$^{17}$ Department of Physics, Sejong University, Seoul, Republic of Korea\\
$^{18}$ Department of Physics, University of California, Berkeley, California, United States\\
$^{19}$ Department of Physics, University of Oslo, Oslo, Norway\\
$^{20}$ Department of Physics and Technology, University of Bergen, Bergen, Norway\\
$^{21}$ Dipartimento di Fisica, Universit\`{a} di Pavia, Pavia, Italy\\
$^{22}$ Dipartimento di Fisica dell'Universit\`{a} and Sezione INFN, Cagliari, Italy\\
$^{23}$ Dipartimento di Fisica dell'Universit\`{a} and Sezione INFN, Trieste, Italy\\
$^{24}$ Dipartimento di Fisica dell'Universit\`{a} and Sezione INFN, Turin, Italy\\
$^{25}$ Dipartimento di Fisica e Astronomia dell'Universit\`{a} and Sezione INFN, Bologna, Italy\\
$^{26}$ Dipartimento di Fisica e Astronomia dell'Universit\`{a} and Sezione INFN, Catania, Italy\\
$^{27}$ Dipartimento di Fisica e Astronomia dell'Universit\`{a} and Sezione INFN, Padova, Italy\\
$^{28}$ Dipartimento di Fisica `E.R.~Caianiello' dell'Universit\`{a} and Gruppo Collegato INFN, Salerno, Italy\\
$^{29}$ Dipartimento DISAT del Politecnico and Sezione INFN, Turin, Italy\\
$^{30}$ Dipartimento di Scienze MIFT, Universit\`{a} di Messina, Messina, Italy\\
$^{31}$ Dipartimento Interateneo di Fisica `M.~Merlin' and Sezione INFN, Bari, Italy\\
$^{32}$ European Organization for Nuclear Research (CERN), Geneva, Switzerland\\
$^{33}$ Faculty of Electrical Engineering, Mechanical Engineering and Naval Architecture, University of Split, Split, Croatia\\
$^{34}$ Faculty of Engineering and Science, Western Norway University of Applied Sciences, Bergen, Norway\\
$^{35}$ Faculty of Nuclear Sciences and Physical Engineering, Czech Technical University in Prague, Prague, Czech Republic\\
$^{36}$ Faculty of Physics, Sofia University, Sofia, Bulgaria\\
$^{37}$ Faculty of Science, P.J.~\v{S}af\'{a}rik University, Ko\v{s}ice, Slovak Republic\\
$^{38}$ Frankfurt Institute for Advanced Studies, Johann Wolfgang Goethe-Universit\"{a}t Frankfurt, Frankfurt, Germany\\
$^{39}$ Fudan University, Shanghai, China\\
$^{40}$ Gangneung-Wonju National University, Gangneung, Republic of Korea\\
$^{41}$ Gauhati University, Department of Physics, Guwahati, India\\
$^{42}$ Helmholtz-Institut f\"{u}r Strahlen- und Kernphysik, Rheinische Friedrich-Wilhelms-Universit\"{a}t Bonn, Bonn, Germany\\
$^{43}$ Helsinki Institute of Physics (HIP), Helsinki, Finland\\
$^{44}$ High Energy Physics Group,  Universidad Aut\'{o}noma de Puebla, Puebla, Mexico\\
$^{45}$ Horia Hulubei National Institute of Physics and Nuclear Engineering, Bucharest, Romania\\
$^{46}$ HUN-REN Wigner Research Centre for Physics, Budapest, Hungary\\
$^{47}$ Indian Institute of Technology Bombay (IIT), Mumbai, India\\
$^{48}$ Indian Institute of Technology Indore, Indore, India\\
$^{49}$ INFN, Laboratori Nazionali di Frascati, Frascati, Italy\\
$^{50}$ INFN, Sezione di Bari, Bari, Italy\\
$^{51}$ INFN, Sezione di Bologna, Bologna, Italy\\
$^{52}$ INFN, Sezione di Cagliari, Cagliari, Italy\\
$^{53}$ INFN, Sezione di Catania, Catania, Italy\\
$^{54}$ INFN, Sezione di Padova, Padova, Italy\\
$^{55}$ INFN, Sezione di Pavia, Pavia, Italy\\
$^{56}$ INFN, Sezione di Torino, Turin, Italy\\
$^{57}$ INFN, Sezione di Trieste, Trieste, Italy\\
$^{58}$ Inha University, Incheon, Republic of Korea\\
$^{59}$ Institute for Gravitational and Subatomic Physics (GRASP), Utrecht University/Nikhef, Utrecht, Netherlands\\
$^{60}$ Institute of Experimental Physics, Slovak Academy of Sciences, Ko\v{s}ice, Slovak Republic\\
$^{61}$ Institute of Physics, Homi Bhabha National Institute, Bhubaneswar, India\\
$^{62}$ Institute of Physics of the Czech Academy of Sciences, Prague, Czech Republic\\
$^{63}$ Institute of Space Science (ISS), Bucharest, Romania\\
$^{64}$ Institut f\"{u}r Kernphysik, Johann Wolfgang Goethe-Universit\"{a}t Frankfurt, Frankfurt, Germany\\
$^{65}$ Instituto de Ciencias Nucleares, Universidad Nacional Aut\'{o}noma de M\'{e}xico, Mexico City, Mexico\\
$^{66}$ Instituto de F\'{i}sica, Universidade Federal do Rio Grande do Sul (UFRGS), Porto Alegre, Brazil\\
$^{67}$ Instituto de F\'{\i}sica, Universidad Nacional Aut\'{o}noma de M\'{e}xico, Mexico City, Mexico\\
$^{68}$ iThemba LABS, National Research Foundation, Somerset West, South Africa\\
$^{69}$ Jeonbuk National University, Jeonju, Republic of Korea\\
$^{70}$ Johann-Wolfgang-Goethe Universit\"{a}t Frankfurt Institut f\"{u}r Informatik, Fachbereich Informatik und Mathematik, Frankfurt, Germany\\
$^{71}$ Korea Institute of Science and Technology Information, Daejeon, Republic of Korea\\
$^{72}$ KTO Karatay University, Konya, Turkey\\
$^{73}$ Laboratoire de Physique Subatomique et de Cosmologie, Universit\'{e} Grenoble-Alpes, CNRS-IN2P3, Grenoble, France\\
$^{74}$ Lawrence Berkeley National Laboratory, Berkeley, California, United States\\
$^{75}$ Lund University Department of Physics, Division of Particle Physics, Lund, Sweden\\
$^{76}$ Nagasaki Institute of Applied Science, Nagasaki, Japan\\
$^{77}$ Nara Women{'}s University (NWU), Nara, Japan\\
$^{78}$ National and Kapodistrian University of Athens, School of Science, Department of Physics , Athens, Greece\\
$^{79}$ National Centre for Nuclear Research, Warsaw, Poland\\
$^{80}$ National Institute of Science Education and Research, Homi Bhabha National Institute, Jatni, India\\
$^{81}$ National Nuclear Research Center, Baku, Azerbaijan\\
$^{82}$ National Research and Innovation Agency - BRIN, Jakarta, Indonesia\\
$^{83}$ Niels Bohr Institute, University of Copenhagen, Copenhagen, Denmark\\
$^{84}$ Nikhef, National institute for subatomic physics, Amsterdam, Netherlands\\
$^{85}$ Nuclear Physics Group, STFC Daresbury Laboratory, Daresbury, United Kingdom\\
$^{86}$ Nuclear Physics Institute of the Czech Academy of Sciences, Husinec-\v{R}e\v{z}, Czech Republic\\
$^{87}$ Oak Ridge National Laboratory, Oak Ridge, Tennessee, United States\\
$^{88}$ Ohio State University, Columbus, Ohio, United States\\
$^{89}$ Physics department, Faculty of science, University of Zagreb, Zagreb, Croatia\\
$^{90}$ Physics Department, Panjab University, Chandigarh, India\\
$^{91}$ Physics Department, University of Jammu, Jammu, India\\
$^{92}$ Physics Program and International Institute for Sustainability with Knotted Chiral Meta Matter (SKCM2), Hiroshima University, Hiroshima, Japan\\
$^{93}$ Physikalisches Institut, Eberhard-Karls-Universit\"{a}t T\"{u}bingen, T\"{u}bingen, Germany\\
$^{94}$ Physikalisches Institut, Ruprecht-Karls-Universit\"{a}t Heidelberg, Heidelberg, Germany\\
$^{95}$ Physik Department, Technische Universit\"{a}t M\"{u}nchen, Munich, Germany\\
$^{96}$ Politecnico di Bari and Sezione INFN, Bari, Italy\\
$^{97}$ Research Division and ExtreMe Matter Institute EMMI, GSI Helmholtzzentrum f\"ur Schwerionenforschung GmbH, Darmstadt, Germany\\
$^{98}$ Saga University, Saga, Japan\\
$^{99}$ Saha Institute of Nuclear Physics, Homi Bhabha National Institute, Kolkata, India\\
$^{100}$ School of Physics and Astronomy, University of Birmingham, Birmingham, United Kingdom\\
$^{101}$ Secci\'{o}n F\'{\i}sica, Departamento de Ciencias, Pontificia Universidad Cat\'{o}lica del Per\'{u}, Lima, Peru\\
$^{102}$ Stefan Meyer Institut f\"{u}r Subatomare Physik (SMI), Vienna, Austria\\
$^{103}$ SUBATECH, IMT Atlantique, Nantes Universit\'{e}, CNRS-IN2P3, Nantes, France\\
$^{104}$ Sungkyunkwan University, Suwon City, Republic of Korea\\
$^{105}$ Suranaree University of Technology, Nakhon Ratchasima, Thailand\\
$^{106}$ Technical University of Ko\v{s}ice, Ko\v{s}ice, Slovak Republic\\
$^{107}$ The Henryk Niewodniczanski Institute of Nuclear Physics, Polish Academy of Sciences, Cracow, Poland\\
$^{108}$ The University of Texas at Austin, Austin, Texas, United States\\
$^{109}$ Universidad Aut\'{o}noma de Sinaloa, Culiac\'{a}n, Mexico\\
$^{110}$ Universidade de S\~{a}o Paulo (USP), S\~{a}o Paulo, Brazil\\
$^{111}$ Universidade Estadual de Campinas (UNICAMP), Campinas, Brazil\\
$^{112}$ Universidade Federal do ABC, Santo Andre, Brazil\\
$^{113}$ Universitatea Nationala de Stiinta si Tehnologie Politehnica Bucuresti, Bucharest, Romania\\
$^{114}$ University of Cape Town, Cape Town, South Africa\\
$^{115}$ University of Derby, Derby, United Kingdom\\
$^{116}$ University of Houston, Houston, Texas, United States\\
$^{117}$ University of Jyv\"{a}skyl\"{a}, Jyv\"{a}skyl\"{a}, Finland\\
$^{118}$ University of Kansas, Lawrence, Kansas, United States\\
$^{119}$ University of Liverpool, Liverpool, United Kingdom\\
$^{120}$ University of Science and Technology of China, Hefei, China\\
$^{121}$ University of South-Eastern Norway, Kongsberg, Norway\\
$^{122}$ University of Tennessee, Knoxville, Tennessee, United States\\
$^{123}$ University of the Witwatersrand, Johannesburg, South Africa\\
$^{124}$ University of Tokyo, Tokyo, Japan\\
$^{125}$ University of Tsukuba, Tsukuba, Japan\\
$^{126}$ Universit\"{a}t M\"{u}nster, Institut f\"{u}r Kernphysik, M\"{u}nster, Germany\\
$^{127}$ Universit\'{e} Clermont Auvergne, CNRS/IN2P3, LPC, Clermont-Ferrand, France\\
$^{128}$ Universit\'{e} de Lyon, CNRS/IN2P3, Institut de Physique des 2 Infinis de Lyon, Lyon, France\\
$^{129}$ Universit\'{e} de Strasbourg, CNRS, IPHC UMR 7178, F-67000 Strasbourg, France, Strasbourg, France\\
$^{130}$ Universit\'{e} Paris-Saclay, Centre d'Etudes de Saclay (CEA), IRFU, D\'{e}partment de Physique Nucl\'{e}aire (DPhN), Saclay, France\\
$^{131}$ Universit\'{e}  Paris-Saclay, CNRS/IN2P3, IJCLab, Orsay, France\\
$^{132}$ Universit\`{a} degli Studi di Foggia, Foggia, Italy\\
$^{133}$ Universit\`{a} del Piemonte Orientale, Vercelli, Italy\\
$^{134}$ Universit\`{a} di Brescia, Brescia, Italy\\
$^{135}$ Variable Energy Cyclotron Centre, Homi Bhabha National Institute, Kolkata, India\\
$^{136}$ Warsaw University of Technology, Warsaw, Poland\\
$^{137}$ Wayne State University, Detroit, Michigan, United States\\
$^{138}$ Yale University, New Haven, Connecticut, United States\\
$^{139}$ Yonsei University, Seoul, Republic of Korea\\
$^{140}$  Zentrum  f\"{u}r Technologie und Transfer (ZTT), Worms, Germany\\
$^{141}$ Affiliated with an institute covered by a cooperation agreement with CERN\\
$^{142}$ Affiliated with an international laboratory covered by a cooperation agreement with CERN.\\

\end{flushleft} 